\documentclass[%
 preprint,
nofootinbib,
amsmath,amssymb,
aps,
showkeys,
]{revtex4-2}

\usepackage{graphicx}
\usepackage{dcolumn}
\usepackage{bm}
\usepackage{xcolor}
\usepackage[mathlines]{lineno}

\begin{document}

\title{Gravitational wave turbulence: a multiple time scale approach for quartic wave interactions}

\author{Benoît Gay}
    \email{benoit.gay@lpp.polytechnique.fr}
\author{S\'ebastien Galtier}
    \email{sebastien.galtier@lpp.polytechnique.fr}
\affiliation{
    Laboratoire de Physique des Plasmas, Université Paris-Saclay, CNRS, École Polytechnique, Sorbonne Université, Observatoire de Paris, F-91128 Palaiseau, France
}

\date{\today}

\begin{abstract}
Wave turbulence is by nature a multiple time scale problem for which there is a natural asymptotic closure. The main result of this analytical theory is the kinetic equation that describes the long-time statistical behaviour of such turbulence composed of a set of weakly nonlinear interacting waves. In the case of gravitational waves, it involves four-wave interactions and two invariants, energy and wave action. Although the kinetic equation of gravitational wave turbulence has been published with the Hadad-Zakharov metric, along with their physical properties, the detailed derivation has not been shown. Following the seminal work of Newell \cite{newell_1968} for gravity/surface waves, we present the multiple time scale method, rarely used to derive the kinetic equations, and clarify the underlying assumptions and methodology. This formalism is applied to a wave amplitude equation obtained using an Eulerian approach. It leads to a kinetic equation slightly different from the one originally published, with a wave equation obtained using a Hamiltonian approach; we verify, however, that the two formulations are fully compatible when the number of symmetries used is the same. 
We also show that the exact solutions (Kolmogorov-Zakharov spectra) exhibit the same power laws and cascade directions. 
Furthermore, the use of the multiple time scale method reveals that the system retains the memory of the initially condition up to a certain level (second order) of development in time.
\end{abstract}


\maketitle

\newpage

\section{Introduction}
\label{sec:introduction}

The initial detection of gravitational waves by the LIGO collaboration \cite{ligoandvirgoscientificcollaborations_2016} came nearly a century after their original theoretical prediction by Albert Einstein \cite{einstein_1918}. The observed deformations on Earth are of very small magnitude ($10^{-21}$), allowing for a predominantly linear approach to study their propagation quickly after they were formed \cite{maggiore_2007}. However, when examining the history of the early universe, certain events may have given rise to gravitational waves of higher amplitude. For instance, we can cite the first-order phase transition \cite{krauss_1992, kosowsky_1992, kamionkowski_1994} or the self-ordering of scalar fields \cite{fenu_2009}. Moreover, gravitational waves of significant amplitude are anticipated to have been generated during cosmic inflation \cite{rubakov_1982, guzzetti_2016}, and numerous endeavours are underway to detect them \cite{planckcollaboration_2016,bertone_2023}. 
The recent publication of the NANOGrav's survey \cite{nanogravcollaboration_2023} provides solid evidence for the existence of a gravitational-wave background. While this result is consistent with a population of supermassive black-holes, we cannot exclude the primordial cosmological origin of this phenomenon.

When dealing with gravitational waves of large amplitude, the linear approach becomes inappropriate and the introduction of non-linear effects inevitable. Among the possible methods, wave turbulence offers a valuable set of tools for examining the weakly non-linear regime, where the injection and dissipation of an invariant (typically the energy or the wave action) take place across different scales. The transfer of this invariant from a larger scale to a smaller scale is referred to as a direct cascade, while the transfer from a smaller scale to a larger scale is known as an inverse cascade. Such cascades manifest in a diverse range of natural phenomena, including surface waves \cite{zakharov_1967}, rotating fluids \cite{galtier_2003}, plasma physics \cite{david_2022}, vibrating plates \cite{during_2006}, Bose-Einstein condensate \cite{nazarenko_2006}, optical waves \cite{dyachenko_1992} and gravitational waves \cite{galtier_2017}.

The foundations of the turbulence theory were laid in particular by Kraichnan's direct interaction approximation, which elucidates the conservation of energy in hydrodynamics through triadic wave interactions \cite{kraichnan_1959}. The main developments in wave turbulence came a little later with Benney and Saffman \cite{benney_1966} who introduced a multiple time scale method \cite{poincare_1892} for a problem involving three-wave interactions, and Benney and Newell \cite{benney_1967c,newell_1968} who extended the theory to quartic interactions with an application to gravity (surface) waves. In both cases, the Eulerian approach was used and the main result was the derivation of the kinetic equation that describes the temporal evolution of spectral invariants such as energy. 
Simultaneously, the Soviet school focused on wave turbulence, particularly in plasmas, using a Hamiltonian approach. Employing a distinct method known as the random phase approximation, they derived the kinetic equations \cite{sagdeev_1969, vedenov_1967, zakharov_1965, zakharov_1967a, zakharov_1967b}, but also found the finite flux spectra (called Kolmogorov-Zakharov spectra) as exact stationary solutions \cite{zakharov_1967, kaner_1970}. 

Over the past ten years, progress has been made in the field of non-linear perturbative approaches to general relativity with a significant focus on the anti-de Sitter space. In particular, it has been shown that this space is unstable under arbitrarily small generic perturbations, leading to a diffusion of energy towards higher frequencies and potentially to the formation of a black hole \cite{bizon_2011}. Numerical investigations have confirmed the plausible turbulent nature of spherically symmetric gravitational collapse in asymptotically anti-de Sitter space, with a scalar curvature showing a power law spectrum close to $-5/3$ \cite{deoliveira_2013}. This result was improperly called Kolmogorov or Kolmogorov-Zakharov spectrum, whereas in turbulence each problem has its own power law energy spectrum index ($-5/3$ being the scaling mainly found in hydrodynamics), and the denomination Kolmogorov-Zakharov is only given to a finite flux spectrum (exact) solution of the kinetic equation \cite{nazarenko_2011,galtier_2017,galtier_2022}. Note that the conclusion on the stability study is not universally applicable as other solutions can remain non-linearly stable \cite{dias_2012, dias_2012a}. 
Interestingly, it is for such an anti-de Sitter space that a technique inspired by the renormalization group theory has been introduced to deal with secular terms that naturally appear in a perturbation theory \cite{craps_2014}. This technique for treating secular terms bears some resemblance to the multiple time scale method that we will present in this paper. Other investigations \cite{craps_2015} have revealed the existence of conserved quantities such as energy or wave action, the latter not well known to non-specialists but often found in wave turbulence. For example, in gravitational wave turbulence, these are two conserved quantities that offer further insights into the dynamics \cite{galtier_2017}. 

It is only recently that significant progress has been made in understanding gravitational wave turbulence, primarily spearheaded by Galtier and Nazarenko. After proving that this turbulence necessarily occurs through quartets of interactions, they used a Hamiltonian approach based on the metric proposed by Hadad and Zakharov \cite{hadad_2014}, which provides a concise and consistent set of equations, to derive the kinetic equation and its exact statistical solutions. They identified two distinct cascades: a direct cascade in energy and an inverse cascade in wave action \cite{galtier_2017}. The latter turned out to be an explosive cascade able in principle of reaching the wave number $\mathbf{k} = \mathbf{0}$ in a finite time \cite{galtier_2019}. However, the wave turbulence description fails at scale $k_s$ where turbulence becomes strongly non-linear. Using a phenomenological model, it is found that the wave action spectrum undergoes a change in slope that remains compatible with an explosive inverse cascade. In the presence of  a small scale excitation, a condensate is expected to form rapidly at small wave numbers. This scenario was then proposed to model the primordial universe: the idea is that the growth of the condensate can be linked to the expansion factor of the Robertson-Walker model \cite{galtier_2020}. The explosive growth would then correspond to a phase of inflation. The associated dilution, and the decrease of the nonlinearities, provides a natural mechanism for stopping the expansion, leaving the universe with a fossil spectrum corresponding to the Harrison-Zeldovich spectrum which is compatible (at main order) with the observations \cite{planckcollaboration_2020}. 
As the contribution of energy tends to zero in the small $k$ limit, this turbulent inflation scenario therefore requires negligible energy.
More recently, using direct numerical simulations, the first direct evidence of a dual cascade has been found \cite{galtier_2021}, proving the relevance of the analytical theory of gravitational wave turbulence.

The aim of this paper is to establish the kinetic equation for gravitational wave turbulence using a multiple time scale approach initially proposed for three-wave interactions \cite{benney_1966} and then generalized to quartic interactions \cite{benney_1967c,newell_1968}. This demonstration was not given in the original paper \cite{galtier_2017}. As turbulence is an obscure subject in cosmology, we think it is important to explain in a pedagogical way (the derivation is long and non-trivial) how to derive the kinetic equation.  Section \ref{sec:expansion} will serve as a reminder of the preliminary assumptions required to derive the evolution equation for the fields within the Hadad-Zakharov metric \cite{hadad_2014}. Here, we follow an Eulerian approach, which is new. Subsequently, in section \ref{sec:derivation}, we will describe the general process of deriving the kinetic equation. This a technical section in which many details are given. In section \ref{sec:application}, we will focus on the specific case of gravitational waves, whose symmetries will have been previously identified. Finally, a conclusion is proposed in section \ref{sec:conclusion}. 

\section{Expansion of the Einstein equation}
\label{sec:expansion}

Throughout the entire paper, we adopt a unit system where the speed of light, $c$, is set to 1. Consequently, the dispersion relation for gravitational waves takes the simple form: $\omega_\mathbf{k} = k$, where $k$ represents the magnitude of the wave vector $\mathbf{k}$. While we do not employ the Einstein summation convention, we use the following convention: 
\begin{equation}
    \delta^{\alpha_1 \dots \alpha_p}_{\beta_1 \dots \beta_q}(\mathbf{k}) = \delta\left(\sum_{i=1}^p \mathbf{k}_{\alpha_i} - \sum_{i=1}^q \mathbf{k}_{\beta_i}\right),
\end{equation} 
and 
\begin{equation}
    \Omega^{\alpha_1 \dots \alpha_p}_{\beta_1 \dots \beta_q} = \sum_{i=1}^p s_{\alpha_i} \omega_{\mathbf{k}_{\alpha_i}} - \sum_{i=1}^q s_{\beta_i} \omega_{\mathbf{k}_{\beta_i}},
\end{equation}
where $s = \pm 1$ is the directional polarity and $\left( \alpha_i \right)_{1 \leq i \leq p}$ and $\left( \beta_i \right)_{1 \leq i \leq q}$ are two family of indices. 

This notation allows for concise representation and facilitate the clarity of the discussion. Below, we provide an overview of the wave amplitude calculation as outlined in \cite{galtier_2017}.

\subsection{Hadad--Zakharov metric}

We begin by considering an observer located in the vacuum at a significant distance from the system under study. This observer utilizes a Cartesian coordinate system $(x, y, z) \in \mathbb{R}^3$ and its own proper time $t \in \mathbb{R}_+$ to describe the interactions. Following \cite{hadad_2014}, we assume that the metric is diagonal and independent of the $z$ coordinate. Thus, $\partial_z$ is a Killing field and the metric can be re-written as the following:
\begin{equation}
g_{\mu \nu} = e^{-2\varphi} \begin{pmatrix}
-(1 + \gamma)^2 & 0 & 0 & 0 \\ 
0 & (1 + \beta)^2 & 0 & 0 \\
0 & 0 & (1 + \alpha)^2 & 0 \\
0 & 0 & 0 & e^{4 \varphi} \\
\end{pmatrix},
\end{equation}
where $\alpha, \beta, \gamma$, and $\varphi$ are real functions that depend solely on $t$, $x$, and $y$. It is important to note that this metric does not take into account the two polarizations of gravitational waves. This simplification considerably reduces the complexity of the system under study, while retaining the main ingredients of wave turbulence: waves and nonlinearities.

The vacuum Einstein equation $R_{\mu \nu} = 0$, with $R_{\mu \nu}$ the Ricci tensor of the metric $g_{\mu \nu}$, yields a set of ten equations. However, as demonstrated in \cite{hadad_2014}, three of these equations are trivially zero, and three are redundant with the remaining four equations. Consequently, we can focus on the following four non-linear equations:
\begin{subequations}
\begin{align}
\partial_x \partial_t \alpha = - &2 (1 + \alpha) \partial_t \varphi \partial_x \varphi + \frac{\partial_t \beta}{1 + \beta} \partial_x \alpha +  \frac{\partial_x \gamma}{1 + \gamma} \partial_t \alpha, \\
\partial_y \partial_t \beta = - &2 (1 + \beta) \partial_t \varphi \partial_y \varphi + \frac{\partial_t \alpha}{1 + \alpha} \partial_y \beta + \frac{\partial_y \gamma}{1 + \gamma} \partial_t \beta, \\
\partial_x \partial_y \gamma = - &2 (1 + \gamma) \partial_x \varphi \partial_y \varphi + \frac{\partial_x \alpha}{1 + \alpha} \partial_y \gamma +  \frac{\partial_y \beta}{1 + \beta} \partial_x \gamma,
\end{align}
and
\begin{equation}
\partial_t \left[ \frac{(1 + \alpha)(1 + \beta)}{1 + \gamma} \partial_t{\varphi} \right] = \partial_x\left[ \frac{(1 + \alpha)(1 + \gamma)}{(1 + \beta)} \partial_x \varphi \right] + \partial_y \left[ \frac{(1 + \beta) (1 + \gamma)}{(1 + \alpha)} \partial_y \varphi \right].
\end{equation}
\label{eq:systeme_initial}
\end{subequations}
The first three expressions correspond to three constraint equations, while the last one describes the non-linear dynamics of the metric. 

In the linear case, the fields $\alpha$, $\beta$, and $\gamma$ are assumed to be null (this is also our initial condition at $t=0$) and consequently only the following two-dimensional wave equation remains:
\begin{equation}
\partial_t \partial_t \varphi - \partial_x \partial_x \varphi - \partial_y \partial_y \varphi = 0.
\label{eq:onde_phi}
\end{equation}
The solutions of (\ref{eq:onde_phi}) are expressed as a superposition of plane waves:
\begin{equation}
\varphi(\mathbf{x}, t) = \int_{\mathbb{R}^2} \left[ A^+(\mathbf{k}) e^{-i(\omega_{\mathbf{k}} t - \mathbf{k} \cdot \mathbf{x})} + A^-(\mathbf{k}) e^{-i(\omega_{\mathbf{k}} t + \mathbf{k} \cdot \mathbf{x})} \right] ~\mathrm{d}^2\mathbf{k},
\label{eq:gal_lambda}
\end{equation}
where by definition $\mathbf{x} = \left( x, y \right)$, $\mathbf{k} = \left( p, q \right)$ and $A^\pm(\mathbf{k})$ are square-integrable functions mapping $\mathbb{R}^2$ to $\mathbb{R}^2$.

To focus on small oscillations, we introduce a small parameter $\epsilon$ (with $0 < \epsilon \ll 1$) which measures the amplitude of the field $\varphi$. The $\alpha$, $\beta$, and $\gamma$ fields can be expanded around their equilibrium positions. At first nonlinear order, equations (\ref{eq:systeme_initial}) can be simplified as follows:
\begin{subequations}
\begin{align}
\partial_x \partial_t \alpha &= - 2 \partial_t \varphi \partial_x \varphi, \label{eq:interest_system_a}\\
\partial_y \partial_t \beta  &= - 2 \partial_t \varphi \partial_y \varphi, \label{eq:interest_system_b}\\
\partial_x \partial_y \gamma &= - 2 \partial_x \varphi \partial_y \varphi,
\label{eq:interest_system_c}
\end{align}
and
\begin{equation}
\partial_t \left[ \left( 1 + {\alpha} + {\beta} - {\gamma} \right) \partial_t \varphi \right] =
\partial_x \left[ \left( 1 + {\alpha} - {\beta} + {\gamma} \right) 
\partial_x \varphi \right] + \partial_y \left[ \left( 1 - {\alpha} + {\beta} + {\gamma} \right) \partial_y \varphi \right].
\label{eq:interest_system_d}
\end{equation}
\label{eq:interest_system}
\end{subequations}
At this stage, the constraint equations (\ref{eq:interest_system_a}--\ref{eq:interest_system_c}) clearly show that the $\alpha$, $\beta$, and $\gamma$ fields depend quadratically on $\varphi$. Therefore, the non-linearities involved in the dynamic equation (\ref{eq:interest_system_d}) are cubic in nature so that we have to deal with four-wave interactions in Fourier space. 

Note that we consider a continuous medium which can lead to mathematical difficulties connected with infinite dimensional phase spaces. For this reason, it is preferable to assume a variable spatially periodic over a box of finite size $L$. However, in the derivation of the kinetic equation, the limit $L \to + \infty$ is finally taken (before the long time limit, or equivalently the limit $\epsilon \to 0$). As both approach leads to the same kinetic equation, for simplicity, we anticipate this result and follow the original approach of \cite{benney_1966}.

\subsection{Wave amplitude equation}

Now, we introduce the normal variables $a^s(\mathbf{k}, t)$ defined as follows:
\begin{equation} \label{eq2p6}
a^s(\mathbf{k}, t) = \frac{1}{\epsilon} \left( \sqrt{\frac{k}{2}} \hat \varphi(\mathbf{k}, t) + \frac{i s}{\sqrt{2 k}} \partial_t \hat \varphi(\mathbf{k}, t) \right) e^{i s \omega_\mathbf{k} t},
\end{equation}
with $s = \pm 1$ the directional polarity (for a given $\mathbf{k}$ we have two directions of propagation) and $\hat \varphi(\mathbf{k}, t)$ the spatial Fourier transform of $\varphi(\mathbf{x}, t)$. 
Conversely, we have the relations:
\begin{subequations}
\begin{equation}
    \hat \varphi(\mathbf{k}, t) = \frac{\epsilon}{\sqrt{2k}} \sum_s a^s_k e^{-i s \omega_\mathbf{k} t},
\end{equation}
and
\begin{equation}
    \partial_t \hat \varphi(\mathbf{k}, t) = -i \epsilon \sqrt{\frac{k}{2}} \sum_s s a^s_k e^{-i s \omega_\mathbf{k} t}.
\end{equation}
\end{subequations}
Referring to the supplementary material of \cite{galtier_2017}, we obtain explicit expressions for $\alpha(\mathbf{k}, t)$, $\beta(\mathbf{k}, t)$, and $\gamma(\mathbf{k}, t)$ in terms of the normal variables using equations (\ref{eq:interest_system_a}), (\ref{eq:interest_system_b}), and (\ref{eq:interest_system_c}). In this context, we assume that the amplitude evolves much slower than the phase, namely:
\begin{equation}
	\frac{\partial_t a^s(\mathbf{k})}{a^s(\mathbf{k})} \ll \omega_{\boldsymbol{k}}.
\end{equation}
This assumption, which is fully consistent with the multiple timescale method introduced in the next section, allows us to treat the amplitude as approximately constant compared to the phase variation during a time integration. Additionally, we consider the integration variables as “dumb”, enabling us to symmetrize the integrands.

The Fourier transform of equations (\ref{eq:interest_system_a})--(\ref{eq:interest_system_b}) leads to the following expressions:
\begin{equation}
	\dot{\alpha}(\mathbf{k})
	= i \frac{\epsilon^2}{2} \int_{\mathbb{R}^4} \sum_{s_1, s_2}
	\frac{s_1 k_1 p_2 + s_2 k_2 p_1}{p \sqrt{k_1 k_2}} 
	a^{s_1}(\mathbf{k}_1) a^{s_2}(\mathbf{k}_2)
    e^{i \Omega_{1 2} t} \delta^0_{12}(\mathbf{k})
	\mathrm{d}^2\mathbf{k}_1 \mathrm{d}^2\mathbf{k}_2,
	\label{eq:alphadot}
\end{equation}
and
\begin{equation}
	\dot{\beta}(\mathbf{k}) 
	= i \frac{\epsilon^2}{2} \int_{\mathbb{R}^4} \sum_{s_1, s_2} 
	\frac{s_1 k_1 q_2 + s_2 k_2 q_1}{q \sqrt{k_1 k_2}} 
	a^{s_1}(\mathbf{k}_1) a^{s_2}(\mathbf{k}_2) e^{i \Omega_{1 2} t} \delta^0_{12}(\mathbf{k})
	\mathrm{d}^2\mathbf{k}_1 \mathrm{d}^2\mathbf{k}_2.
	\label{eq:betadot}
\end{equation}
These equations can be integrated over time, resulting in the following expressions:
\begin{equation}
	\alpha(\mathbf{k}) =
	- \frac{\epsilon^2}{2} \int_{\mathbb{R}^4} \sum_{s_1, s_2} 
	\frac{s_1 k_1 p_2 + s_2 k_2 p_1}{p \sqrt{k_1 k_2}}
	\frac{a^{s_1}(\mathbf{k}_1) a^{s_2}(\mathbf{k}_2)}{s_1 k_1 + s_2 k_2} e^{i \Omega_{1 2} t} \delta^0_{12}(\mathbf{k})
	\mathrm{d}^2\mathbf{k}_1 \mathrm{d}^2\mathbf{k}_2
	\label{eq:alpha}
\end{equation}
and,
\begin{equation}
	\beta(\mathbf{k}) =
	- \frac{\epsilon^2}{2} \int_{\mathbb{R}^4} \sum_{s_1, s_2} 
	\frac{s_1 k_1 q_2 + s_2 k_2 q_1}{q \sqrt{k_1 k_2}}
	\frac{a^{s_1}(\mathbf{k}_1) a^{s_2}(\mathbf{k}_2)}{s_1 k_1 + s_2 k_2} 
    e^{i \Omega_{1 2} t} \delta^0_{12}(\mathbf{k})
	\mathrm{d}^2\mathbf{k}_1 \mathrm{d}^2\mathbf{k}_2.
	\label{eq:beta}
\end{equation}
In addition, the Fourier transform of equation (\ref{eq:interest_system_c}) leads to the following expression:
\begin{equation}
	\gamma(\mathbf{k}) =
	- \frac{\epsilon^2}{2} \int_{\mathbb{R}^4} \sum_{s_1, s_2} 
	\frac{p_1 q_2 + p_2 q_1}{pq} \frac{1}{\sqrt{k_1 k_2}}
	a^{s_1}(\mathbf{k}_1) a^{s_2}(\mathbf{k}_2)
    e^{i \Omega_{1 2} t} \delta^0_{12}(\mathbf{k})
	\mathrm{d}^2\mathbf{k}_1 \mathrm{d}^2\mathbf{k}_2.
	\label{eq:gamma}
\end{equation}
This expression can be differentiated with respect to time, resulting in:
\begin{equation}
	\dot{\gamma}(\mathbf{k}) =
	i \frac{\epsilon^2}{2} \int_{\mathbb{R}^4} \sum_{s_1, s_2} 
	\frac{p_1 q_2 + p_2 q_1}{pq} \frac{s_1 k_1 + s_2 k_2}{\sqrt{k_1 k_2}}
    a^{s_1}(\mathbf{k}_1) a^{s_2}(\mathbf{k}_2)
    e^{i \Omega_{1 2} t} \delta^0_{12}(\mathbf{k})
	\mathrm{d}^2\mathbf{k}_1 \mathrm{d}^2\mathbf{k}_2.
	\label{eq:gammadot}
\end{equation}

Next, we substitute all the previously derived expressions into equation (\ref{eq:interest_system_d}) in order to find the wave amplitude equation. We obtain after manipulations:
\begin{equation}
	\partial_t a^s(\mathbf{k}) = 
	\epsilon^2 
	\int_{\mathbb{R}^6} \sum_{s_1, s_2, s_3}
	\mathbf{L}^{-s s_1 s_2 s_3}_{-\mathbf{k} \mathbf{k}_1 \mathbf{k}_2 \mathbf{k}_3}
	a^{s_1}(\mathbf{k}_1) a^{s_2}(\mathbf{k}_2) a^{s_3}(\mathbf{k}_3) e^{i \Omega^{0}_{1 2 3} t} ~\delta^0_{123}(\mathbf{k}) ~\prod_{i=1}^3 \mathrm{d}^2\mathbf{k}_i ,
	\label{eq:evolution_amplitude}
\end{equation}
where $\mathbf{L}^{-s s_1 s_2 s_3}_{-\mathbf{k} \mathbf{k}_1 \mathbf{k}_2 \mathbf{k}_3}$ is the interaction coefficient, defined as:
\begin{equation}
	\mathbf{L}^{s s_1 s_2 s_3}_{\mathbf{k} \mathbf{k}_1 \mathbf{k}_2 \mathbf{k}_3} 
	= \frac{is}{12 \sqrt{k k_1 k_2 k_3}} \left( 
	\mathbf{T}^{s s_1 s_2 s_3}_{\mathbf{k} \mathbf{k}_1 \mathbf{k}_2 \mathbf{k}_3} + 
	\mathbf{T}^{s s_2 s_3 s_1}_{\mathbf{k} \mathbf{k}_2 \mathbf{k}_3 \mathbf{k}_1} + 
	\mathbf{T}^{s s_3 s_2 s_1}_{\mathbf{k} \mathbf{k}_3 \mathbf{k}_2 \mathbf{k}_1} \right),
	\label{eq:coefficient_interaction}
\end{equation}
with:
\begin{equation}
\begin{aligned}
	\mathbf{T}^{s s_1 s_2 s_3}_{\mathbf{k} \mathbf{k}_1 \mathbf{k}_2 \mathbf{k}_3} =&
	\bigg[
    \left( p_2 q_3 + p_3 q_2 \right) \frac{ p p_1 + q q_1 - s_1 k_1 \left(s_1 k_1 + s_2 k_2 + s_3 k_3\right)}{\left(p_2 + p_3\right) \left(q_2 + q_3\right)} \\
	& \hspace{5pt} 
    + \frac{s_2 k_2 p_3 + s_3 k_3 p_2}{s_2 k_2 + s_3 k_3} \frac{p p_1 - q q_1 + s_1 k_1 \left(s_1 k_1 + s_2 k_2 + s_3 k_3\right)}{p_2 + p_3} \\
    & \hspace{5pt} 
    - \frac{s_2 k_2 q_3 + s_3 k_3 q_2}{s_2 k_2 + s_3 k_3} \frac{p p_1 - q q_1 - s_1 k_1 \left(s_1 k_1 + s_2 k_2 + s_3 k_3\right)}{q_2 + q_3} 
    \bigg].
\end{aligned}
\end{equation}

\subsection{Symmetries of the interaction coefficient}

The specific form of the interaction coefficient is determined by considering the dummy character of the integration variable, which allows us to identify certain symmetries of the coefficient. In general, the interaction coefficient exhibits the following properties:
\begin{itemize}
\begin{subequations}
    \item Homogeneity: It is a homogeneous function of the wave vectors $\mathbf{k}_i$. This means that there exists a constant $b \in \mathbb{R}$ such that, for all $a \in \mathbb{R}$,
    \begin{equation}
        \mathbf{L}^{s s_1 s_2 s_3}_{a\mathbf{k} a\mathbf{k}_1 a\mathbf{k}_2 a\mathbf{k}_3} = a^b \mathbf{L}^{s s_1 s_2 s_3}_{\mathbf{k} \mathbf{k}_1 \mathbf{k}_2 \mathbf{k}_3}.
    \end{equation}
    It is worth noting that, for our specific problem, $b$ is equal to zero.
    \item Imaginary nature: The interaction coefficient is purely imaginary, satisfying the relation:
    \begin{equation}
        \left(\mathbf{L}^{s s_1 s_2 s_3}_{\mathbf{k} \mathbf{k}_1 \mathbf{k}_2 \mathbf{k}_3}\right)^* = - \mathbf{L}^{s s_1 s_2 s_3}_{\mathbf{k} \mathbf{k}_1 \mathbf{k}_2 \mathbf{k}_3}.
    \end{equation}
    \item Symmetry: The interaction coefficient exhibits symmetry under certain transformations. Specifically, it is symmetric when the signs of the directional polarities or all wave vectors are changed:
    \begin{equation}
        \mathbf{L}^{- s -s_1 -s_2 -s_3}_{-\mathbf{k} -\mathbf{k}_1 -\mathbf{k}_2 -\mathbf{k}_3} = - \mathbf{L}^{s s_1 s_2 s_3}_{\mathbf{k} \mathbf{k}_1 \mathbf{k}_2 \mathbf{k}_3}.
    \end{equation}
    \item Symmetries on the resonant manifold: On the resonant manifold, where $s \omega_\mathbf{k} = s_1 \omega_{\mathbf{k}_1} + s_2 \omega_{\mathbf{k}_2} + s_3 \omega_{\mathbf{k}_3}$ and $\mathbf{k} = \mathbf{k}_1 + \mathbf{k}_2 + \mathbf{k}_3$, the interaction coefficient exhibits symmetry with respect to the last indices. Thus, for any permutation $\sigma$ of $\left\{1, 2, 3 \right\}$, we have:
    \begin{equation}
    \mathbf{L}^{s s_{\sigma(1)} s_{\sigma(2)} s_{\sigma(3)}}_{\mathbf{k} \mathbf{k}{\sigma(1)} \mathbf{k}{\sigma(2)} \mathbf{k}{\sigma(3)}} = \mathbf{L}^{s s_1 s_2 s_3}_{\mathbf{k} \mathbf{k}_1 \mathbf{k}_2 \mathbf{k}_3}.
    \end{equation}
    Additionally, the interaction coefficient vanishes when $\mathbf{k} = 0$, ensuring the conservation of the null first-order cumulant:
    \begin{equation}
        \mathbf{L}^{s s_1 s_2 s_3}_{\mathbf{0} \mathbf{k}_1 \mathbf{k}_2 \mathbf{k}_3} = 0.
    \end{equation}
\end{subequations}
\end{itemize}

\section{Sequential time closure for quartic-wave interactions}
\label{sec:derivation}

\subsection{A multiple time scale method}

Our statistical study begins with the fundamental equation (\ref{eq:evolution_amplitude}) from which we apply the method of multiple time scales to derive the kinetic equation. Note that this method was originally developed for three-wave interactions \cite{benney_1966} and then to quartic interactions with gravity waves as its main application \cite{benney_1967c, newell_1968}. The original problem (surface waves) is 2-dimensional, however, the method is valid in a space of n-dimension, with $n \geq 1$. That is why, for this section only, we will perform the spatial integrations on vectors of $\mathbb{R}^n$.

We introduce a set of (approximately) independent variables: 
\begin{equation} 
T_0 = t, \quad T_1 = \epsilon t, \quad T_2 = \epsilon^2 t,  \quad \dots
\end{equation}
with $\epsilon$ a small parameter ($0 < \epsilon \ll 1$) which measures the wave amplitude (see section \ref{sec:expansion}). We apply the chain rule, yielding:
\begin{equation}
\partial_t = \sum_{i = 0}^{+\infty} \epsilon^i \partial_{T_i} = \partial_{T_0} + \epsilon \partial_{T_1} + \epsilon^2 \partial_{T_2} + \dots
\label{eq:decomposition_t}
\end{equation}
The normal variables are also expanded to the power of $\epsilon$ as follows:
\begin{equation}
a^{s}(\mathbf{k}) = \sum_{i = 0}^{+\infty} \epsilon^i a^{s}_i(\mathbf{k}) = a^{s}_0(\mathbf{k}) + \epsilon a^{s}_1(\mathbf{k}) + \epsilon^2 a^{s}_2(\mathbf{k}) + \dots
\label{eq:decomposition_vc}
\end{equation}
In order to enhance readability, we disregard the time dependence, so $a^s(\mathbf{k})$ must be understood like $a^{s}(\mathbf{k}, T_0, T_1, T_2, \dots)$. Furthermore, we assume that the phase of the normal variables only varies on the $T_0$ timescale. The justification for this assumption will be provided later.

With these definitions, we obtain the following expansion:
\begin{multline}
\sum_{i,j = 0}^{+\infty} \epsilon^{i+j} \partial_{T_i} a^{s}_j(\mathbf{k}) = \\
\sum_{p,q,r = 0}^{+\infty} \epsilon^{p+q+r+2} \int_{\mathbb{R}^{3n}} \sum_{s_1, s_2, s_3} \mathbf{L}^{-s s_1 s_2 s_3}_{-\mathbf{k} \mathbf{k}_1 \mathbf{k}_2 \mathbf{k}_3} a^{s_1}_p(\mathbf{k}_1) a^{s_2}_q(\mathbf{k}_2) a^{s_3}_r(\mathbf{k}_3) ~\delta^0_{123}(\mathbf{k}) ~e^{i \Omega^{0}_{1 2 3} T_0} ~\prod_{i=1}^3 \mathrm{d}^n\mathbf{k}_i. 
\label{eq:amplitude_equation_developed_compacted}  
\end{multline}
This expression leads to a hierarchy of equations in different orders in $\epsilon$. Using eq. (\ref{eq:amplitude_equation_developed_compacted}) to identify the different powers of $\epsilon$, we obtain the following expressions:
\begin{subequations}
\allowdisplaybreaks
\begin{align}
\partial_{T_0} a^{s}_0(\mathbf{k}) =& ~0; \label{eq:dT0a0} \\
\partial_{T_0} a^{s}_1(\mathbf{k}) =& -\partial_{T_1} a^{s}_0(\mathbf{k}); \label{eq:dT0a1} \\
\partial_{T_0} a^{s}_2(\mathbf{k}) =& -\partial_{T_2} a^{s}_0(\mathbf{k}) - \partial_{T_1} a^{s}_1(\mathbf{k}) \\
&+ \int_{\mathbb{R}^{3n}} \sum_{s_1, s_2, s_3} \mathbf{L}^{-s s_1 s_2 s_3}_{-\mathbf{k} \mathbf{k}_1 \mathbf{k}_2 \mathbf{k}_3} a^{s_1}_0(\mathbf{k}_1) a^{s_2}_0(\mathbf{k}_2) a^{s_3}_0(\mathbf{k}_3) \delta^{0}_{1 2 3}(\mathbf{k}) e^{i \Omega^0_{1 2 3} T_0} \prod_{i=1}^3 \mathrm{d}^n\mathbf{k}_i; \label{eq:dT0a2} \\
\partial_{T_0} a^{s}_3(\mathbf{k}) =& -\partial_{T_3} a^{s}_0(\mathbf{k}) - \partial_{T_2} a^{s}_1(\mathbf{k}) - \partial_{T_1} a^{s}_2(\mathbf{k}) \\
&+ 3 \int_{\mathbb{R}^{3n}} \sum_{s_1, s_2, s_3} \mathbf{L}^{-s s_1 s_2 s_3}_{-\mathbf{k} \mathbf{k}_1 \mathbf{k}_2 \mathbf{k}_3} a^{s_1}_1(\mathbf{k}_1) a^{s_2}_0(\mathbf{k}_2) a^{s_3}_0(\mathbf{k}_3) \delta^{0}_{1 2 3}(\mathbf{k}) e^{i \Omega^0_{1 2 3} T_0} \prod_{i=1}^3 \mathrm{d}^n\mathbf{k}_i;\label{eq:dT0a3} \\
\begin{split}
\partial_{T_0} a^{s}_4(\mathbf{k}) =& -\partial_{T_4} a^{s}_0(\mathbf{k}) - \partial_{T_3} a^{s}_1(\mathbf{k}) - \partial_{T_2} a^{s}_2(\mathbf{k}) - \partial_{T_1} a^{s}_3(\mathbf{k}) \\
&+ 3 \int_{\mathbb{R}^{3n}} \sum_{s_1, s_2, s_3} \mathbf{L}^{-s s_1 s_2 s_3}_{-\mathbf{k} \mathbf{k}_1 \mathbf{k}_2 \mathbf{k}_3} a^{s_1}_2(\mathbf{k}_1) a^{s_2}_0(\mathbf{k}_2) a^{s_3}_0(\mathbf{k}_3) \delta^{0}_{1 2 3}(\mathbf{k}) e^{i \Omega^0_{1 2 3} T_0} \prod_{i=1}^3 \mathrm{d}^n\mathbf{k}_i \\
&+ 3 \int_{\mathbb{R}^{3n}} \sum_{s_1, s_2, s_3} \mathbf{L}^{-s s_1 s_2 s_3}_{-\mathbf{k} \mathbf{k}_1 \mathbf{k}_2 \mathbf{k}_3} a^{s_1}_1(\mathbf{k}_1) a^{s_2}_1(\mathbf{k}_2) a^{s_3}_0(\mathbf{k}_3) \delta^{0}_{1 2 3}(\mathbf{k}) e^{i \Omega^0_{1 2 3} T_0} \prod_{i=1}^3 \mathrm{d}^n\mathbf{k}_i.\label{eq:dT0a4}
\end{split}
\end{align}
\end{subequations}
The factor of $3$ arises from considering the symmetries of the interaction coefficient. Additionally, we observe that eq. (\ref{eq:dT0a0}) provides justification for assuming that the amplitude and phase do not vary on the same timescale.

We choose initial conditions (at $t=0$) such that: $\forall i \ge 1$, $a_i^s(\mathbf{k}) = 0$. The integration of the previous equation with respect to $T_0$ gives, 
after a few manipulations:
\begin{subequations}
\begin{align}
a^{s}_0(\mathbf{k}) =&~ a^{s}_0(\mathbf{k}, T_1, T_2, \dots); \label{eq:a0} \\
a^{s}_1(\mathbf{k}) =& - T_0 \partial_{T_1} a^{s}_0(\mathbf{k}); \label{eq:a1} \\
a^{s}_2(\mathbf{k}) =& - T_0 \partial_{T_2} a^{s}_0(\mathbf{k}) + \frac{T_0^{\phantom{a}2}}{2} \partial^2_{T_1} a^{s}_0(\mathbf{k}) + b^{s}_2(\mathbf{k}); \label{eq:a2} \\
a^{s}_3(\mathbf{k}) =& - T_0 \partial_{T_3} a^{s}_0(\mathbf{k}) + T_0^{\phantom{a}2} \partial_{T_1} \partial_{T_2} a^{s}_0(\mathbf{k}) - \frac{T_0^{\phantom{a}3}}{6} \partial^3_{T_1} a^{s}_0(\mathbf{k}) -T_0 \partial_{T_1} b_2^s(\mathbf{k}); \label{eq:a3} \\
\begin{split}
	a^{s}_4(\mathbf{k}) =& - T_0 \partial_{T_4} a^{s}_0(\mathbf{k}) 
	+ \frac{T_0^{\phantom{a}2}}{2} \left[ 2\partial_{T_1} \partial_{T_3} a^{s}_0(\mathbf{k}) + \partial_{T_2}^2 a^{s}_0(\mathbf{k}) \right] - \frac{T_0^{\phantom{a}3}}{2} \partial_{T_1}^2 \partial_{T_2}  a^{s}_0(\mathbf{k}) \\
    & + \frac{T_0^{\phantom{a}4}}{24} \partial_{T_1}^4 a^{s}_0(\mathbf{k}) + b_4^s(\mathbf{k}) - T_0 \partial_{T_2} b_2^s(\mathbf{k}) + \frac{T_0^{\phantom{a}2}}{2} \partial_{T_1}^2 b_2^s (\mathbf{k}); \label{eq:a4}
\end{split}
\end{align}
\end{subequations}
where:
\begin{subequations}
\begin{align}
	\displaystyle b^{s}_2(\mathbf{k}) =&~ \int_{\mathbb{R}^{3n}} \sum_{s_1, s_2, s_3} \mathbf{L}^{-s s_1 s_2 s_3}_{-\mathbf{k} \mathbf{k}_1 \mathbf{k}_2 \mathbf{k}_3} a^{s_1}_0(\mathbf{k}_1) a^{s_2}_0(\mathbf{k}_2) a^{s_3}_0(\mathbf{k}_3) \delta^{0}_{1 2 3}(\mathbf{k}) \Delta\left(\Omega^0_{1 2 3}, T_0\right) \prod_{i=1}^3 \mathrm{d}^n\mathbf{k}_i; \\
	\begin{split}
		b^{s}_4(\mathbf{k}) =& ~3 \int_{\mathbb{R}^{6n}} \sum_{\substack{s_1, s_2, s_3 \\ s_4, s_5, s_6}} \mathbf{L}^{-s s_1 s_2 s_3}_{-\mathbf{k} \mathbf{k}_1 \mathbf{k}_2 \mathbf{k}_3} \mathbf{L}^{-s_1 s_4 s_5 s_6}_{-\mathbf{k}_1 \mathbf{k}_4 \mathbf{k}_5 \mathbf{k}_6} a^{s_2}_0(\mathbf{k}_2) a^{s_3}_0(\mathbf{k}_3) a^{s_4}_0(\mathbf{k}_4) a^{s_5}_0(\mathbf{k}_5) a^{s_6}_0(\mathbf{k}_6) \\[-18pt]
		& \hspace{131pt} \times  \delta^{0}_{1 2 3}(\mathbf{k}) \delta^{1}_{4 5 6}(\mathbf{k}) E\left(\Omega^0_{2 3 4 5 6}, \Omega^0_{1 2 3}, T_0\right) \prod_{i=1}^6 \mathrm{d}^n\mathbf{k}_i;
	\end{split}
\end{align}
\end{subequations}
with:
\begin{equation*}
	\Delta\left(X, T\right) = \frac{e^{iXT} - 1}{iX} \quad \text{and} \quad E\left(X, Y, T\right) = \int_0^T \Delta(X-Y,t) e^{iYt}~\mathrm{d}t.
\end{equation*}

\subsection{Statistical assumptions}

We make the statistical assumption that $\langle a^{s}(\mathbf{k}) \rangle = 0$, where $\langle  \rangle$ is the ensemble average. Going back to general relativity, this situation corresponds to space-time fluctuations around a flat metric. 
We also assume that $\varphi(\mathbf{r})$ and $\partial_t\varphi(\mathbf{r})$ are spatially homogeneous, meaning that we are in the heart of the turbulence. As a result, the corresponding normal variables $a^{s}(\mathbf{k})$ are also spatially homogeneous since they are defined as linear combinations of these fields. The second-order moment can thus be written in terms of the second-order cumulant $q^{s s'}(\mathbf{k},\mathbf{k}')$ as follows:
\begin{equation}
    \langle a^{s}_0(\mathbf{k}) a^{s'}_0(\mathbf{k}') \rangle = q_0^{s s'}(\mathbf{k},\mathbf{k}') \delta\left(\mathbf{k} + \mathbf{k}'\right),
	\label{eq:aa_mean_decomposition}
\end{equation}
where the presence of $\delta\left(\mathbf{k} + \mathbf{k}'\right)$ is the consequence of statistical homogeneity. Note that for the derivation of the kinetic equation, the product $\langle a^{s}(\mathbf{k}) a^{s'}(\mathbf{k}') \rangle$ will finally be treated in the special case where $s' = -s$. The reason is that the quantities of interest are wave action ($\langle \vert a^{s}(\mathbf{k}) \vert^2 \rangle = \langle a^{s}(\mathbf{k}) a^{-s}(\mathbf{-k}) \rangle$) and energy ($\omega_{\mathbf{k}} \langle \left|a^{s}(\mathbf{k})\right|^2 \rangle$) spectra which are real and positive.

The  $n$th-order moment $\langle a^{s_1}(\mathbf{k}_1) \dots a^{s_r}(\mathbf{k}_r) \rangle$ can be written in terms of lower order cumulants $q^{s_1 \dots s_r}(\mathbf{k}_1, \dots, \mathbf{k}_r)$. For instance, if $r=4$, we have for the leading order cumulants:
\begin{equation}
\begin{aligned}
    \langle a^{s_1}_0(\mathbf{k}_1) a^{s_2}_0(\mathbf{k}_2) a^{s_3}_0(\mathbf{k}_3) a^{s_4}_0(\mathbf{k}_4) \rangle =& ~q_0^{s_1 s_2 s_3 s_4}(\mathbf{k}_1, \mathbf{k}_2, \mathbf{k}_3, \mathbf{k}_4) \delta^{1234}(\mathbf{k}) \\
    &+ q_0^{s_1 s_2}(\mathbf{k}_1, \mathbf{k}_2) q_0^{s_3 s_4}(\mathbf{k}_3, \mathbf{k}_4) \delta^{12}(\mathbf{k})  \delta^{34}(\mathbf{k}) \\ 
    &+ q_0^{s_1 s_3}(\mathbf{k}_1, \mathbf{k}_3) q_0^{s_2 s_4}(\mathbf{k}_2, \mathbf{k}_4) \delta^{13}(\mathbf{k})  \delta^{24}(\mathbf{k}) \\
    &+ q_0^{s_1 s_4}(\mathbf{k}_1, \mathbf{k}_4) q_0^{s_2 s_3}(\mathbf{k}_2, \mathbf{k}_3) \delta^{14}(\mathbf{k})  \delta^{23}(\mathbf{k}) .
\end{aligned}
\end{equation}
Note that the useful case $r=6$ is recalled in appendix \ref{sec:appendix_six_points_cummulant}.

To simplify the analysis, we will assume Gaussian \footnote{Note that the initial Gaussian assumption is a priori not required for the derivation of the kinetic equation, and we can also assume, as often \cite{nazarenko_2011}, initial fields $a_0^s(\mathbf{k})$ with random phase and amplitude.} initial distribution for $a_0^s(\mathbf{k})$ so that the fourth-order cumulant at order zero vanishes, $q^{s s' s'' s'''}_0(\mathbf{k}, \mathbf{k}', \mathbf{k}'', \mathbf{k}''', t=0) = 0$. This assumption is sufficient to guarantee the absence of coherent structure in the initial excitation that could invalidate our derivation based on weak nonlinearities.
Under the additional assumption of boundedness of the fourth-order moment when $T_0 \to + \infty$, we will show that this initial property propagates at time $T_2$ and even has an impact on the kinetic equation at time $T_4$. This phenomenon is sometimes referred to as propagation of chaos \cite{deng_2022}.

A fundamental aspect of the multiple timescale method is the assumption that the second-order moment $\langle a^{s}(\mathbf{k}) a^{s'}(\mathbf{k}') \rangle$ remains bounded as $T_0 \to + \infty$. Making use of the previous definitions, we expend this term according to:
\begin{equation}
\begin{aligned}
    \langle a^{s}(\mathbf{k}) a^{s'}(\mathbf{k}') \rangle &= \left\langle \left( a^{s}_0(\mathbf{k}) + \epsilon a^{s}_1(\mathbf{k}) + \epsilon^2 a^{s\phantom{'}}_2(\mathbf{k}) + \dots \right) \right.\\
    & \times \left. \left( a^{s'}_0(\mathbf{k}') + \epsilon a^{s'}_1(\mathbf{k}') + \epsilon^2 a^{s'}_2(\mathbf{k}') + \dots \right) \right\rangle \\
    &= \sum_{n=0}^{+\infty} \epsilon^n \sum_{i=0}^n \langle a^{s}_i(\mathbf{k}) a^{s}_{n-i}(\mathbf{k}') \rangle.
\end{aligned}
\label{eq:aaa}
\end{equation}
Thus, the contribution of the right-hand side must also be bounded at each order $\epsilon^n$, with $n \geq 0$. In practice, we will see that secular terms may appear and, therefore, we will impose the nullity of these contributions at a given order to keep the development uniform in time. It is precisely this condition at order $\epsilon^4$ that will lead us to the kinetic equation. 

Our last statistical assumption concerns the boundedness of the fourth-order cumulants over time. In other words, we impose that $\langle a^{s_1}(\mathbf{k}_1) a^{s_2}(\mathbf{k}_2) a^{s_3}(\mathbf{k}_3) a^{s_4}(\mathbf{k}_4) \rangle$ does not diverge when $T_0 \to + \infty$. Thus, we will have to consider sums of products given by:
\begin{equation}
    \sum_{\substack{i_1, i_2, i_3, i_4 \\ i_1 + i_2 + i_3 + i_4 = n}} \langle a^{s_1}_{i_1}(\mathbf{k}_1) a^{s_2}_{i_2}(\mathbf{k}_2) a^{s_3}_{i_3}(\mathbf{k}_3) a^{s_4}_{i_4}(\mathbf{k}_4) \rangle,
\end{equation}
and impose the nullity of the corresponding secular terms.

\subsection{First asymptotic closure at time $\mathbf{T_1}$}

Now we will derive the conditions which ensure the second-order moment remains bounded at all times. This involves counting all the secular terms at each order $\epsilon^n$, $n \ge 0$ and cancelling their contributions. We will see that it is the condition of order $\epsilon^4$ that gives the kinetic equation we are looking for. 
Using expression (\ref{eq:aaa}), we see that at order $\epsilon^0$ the result is immediate: the only  term is $\langle a^{s}_0(\mathbf{k}) a^{s'}_0(\mathbf{k}') \rangle$ which is bounded, so there is no secular contribution and therefore no asymptotic condition to impose. 

At order $\epsilon$ the situation is different because we have:
\begin{equation}
\langle a^{s}_0(\mathbf{k}) a^{s'}_1(\mathbf{k}') + a^{s}_1(\mathbf{k}) a^{s'}_0(\mathbf{k}') \rangle = - T_0 \partial_{T_1} \langle a^{s}_0(\mathbf{k}) a^{s'}_0(\mathbf{k}') \rangle.
\end{equation}
In order to keep the left-hand side bounded, we must impose the condition:
\begin{equation}
\partial_{T_1} \langle a^{s}_0(\mathbf{k}) a^{s'}_0(\mathbf{k}') \rangle = 0,
\label{eq:dT1wa}
\end{equation}
which means that the wave action does not evolve on the $T_1$ timescale. Therefore, the turbulent cascade is not effective on this short time scale. Similarly, we can justify that:
\begin{equation}
\partial_{T_1} \langle a^{s}_0(\mathbf{k}) a^{s'}_0(\mathbf{k}') a^{s''}_0(\mathbf{k}'') a^{s'''}_0(\mathbf{k}''')\rangle = 0,
\label{eq:dT1wa2}
\end{equation}
based on the bounded nature of the fourth-order moment\footnote{These results based on the boundedness of second and fourth order moments can be assumed for any moments of order $n$. Therefore, the probability distribution does not depend on $T_1$ and we may assume that the $a^{s}_0(\mathbf{k})$ variable itself does not depend on $T_1$ either. Thus, at most we have: $a^{s}_0(\mathbf{k}) = a^{s}_0(\mathbf{k}, T_2, T_3, \dots)$.
This justifies the requirement that the $T_1$ timescale is irrelevant in the derivation of the kinetic equation as it was done by Newell \cite{newell_1968}. Nevertheless, in our derivation, we will keep the $T_1$ dependence to show that this dependence can disappear with this boundedness condition.}

\subsection{Second asymptotic closure at time $\mathbf{T_2}$}

We continue the analysis with the second order:
\begin{multline}
	\langle a^{s}_0(\mathbf{k}) a^{s'}_2(\mathbf{k}') + a^{s}_1(\mathbf{k}) a^{s'}_1(\mathbf{k}') + a^{s}_2(\mathbf{k}) a^{s'}_0(\mathbf{k}') \rangle \\
    = - T_0 \partial_{T_2} \langle a^{s}_0(\mathbf{k}) a^{s'}_0(\mathbf{k}') \rangle + \frac{T_0^{\phantom{a}2}}{2} \partial_{T_1}^2 \langle a^{s}_0(\mathbf{k})  a^{s'}_0(\mathbf{k}') \rangle + \langle a^{s}_0(\mathbf{k}) b^{s'}_2(\mathbf{k}') + a^{s'}_0(\mathbf{k}') b^{s}_2(\mathbf{k}) \rangle.
\end{multline}
This expression can be simplified using the first closure (\ref{eq:dT1wa}). Therefore, to prevent unwanted secular growth of the left-hand side, we must balance the terms proportional to $T_0$:
\begin{equation}
	\partial_{T_2} \langle a^{s}_0(\mathbf{k}) a^{s'}_0(\mathbf{k}') \rangle = \mathcal{C}_{T_0} \langle a^{s}_0(\mathbf{k}) b^{s'}_2(\mathbf{k}') + a^{s'}_0(\mathbf{k}') b^{s}_2(\mathbf{k}) \rangle,
\end{equation}
where $\mathcal{C}_{T_0^{\phantom{a}n}} \mathcal{E}$ refers to the term proportional to $T_0^{\phantom{a}n}$ in $\mathcal{E}$. 
The long time contribution of these oscillating integrals is given by the theory of generalized function and the Riemann-Lebesgue lemma (see Appendix \ref{sec:appendix_RL_for_distributions}):
\begin{equation}
	\Delta(X, T) \xrightarrow[T_0 \to + \infty]{} \begin{cases}
		\pi \delta(X) + i \mathcal{P} \left( \frac{1}{X} \right), & \text{if}~X \neq 0 \\
		T_0, & \text{if}~X = 0 \\
	\end{cases} ;
	\label{eq:lemme_RiemannLebesgue}
\end{equation}
where $\mathcal{P}$ refers to the Cauchy principal value. This means that $\langle a^{s}_0(\mathbf{k}) b^{s'}_2(\mathbf{k}') + a^{s'}_0(\mathbf{k}') b^{s}_2(\mathbf{k}) \rangle$ can exhibit secular growth when the different Dirac deltas constrain the integration to be performed on the resonant manifold. 

We illustrate the computation with an example:
\begin{equation}
\begin{aligned}
	\langle a^{s}_0(\mathbf{k}) b^{s'}_2(\mathbf{k}') \rangle & \\
    &\hspace{-50pt} = \int_{\mathbb{R}^{3n}} \sum_{s_1, s_2, s_3} \mathbf{L}^{-s' s_1 s_2 s_3}_{-\mathbf{k}' \mathbf{k}_1 \mathbf{k}_2 \mathbf{k}_3} \langle a^{s}_0(\mathbf{k})  a^{s_1}_0(\mathbf{k}_1) a^{s_2}_0(\mathbf{k}_2) a^{s_3}_0(\mathbf{k}_3) \rangle \Delta(\Omega^{0'}_{123}, T_0) \delta^{0'}_{1 2 3}(\mathbf{k}) \prod_{i=1}^3 \mathrm{d}^n\mathbf{k}_i \\
	&\hspace{-50pt} = ~3 \int_{\mathbb{R}^{3n}} \sum_{s_1, s_2, s_3} \mathbf{L}^{-s' s_1 s_2 s_3}_{-\mathbf{k}' \mathbf{k}_1 \mathbf{k}_2 \mathbf{k}_3} q_0^{s s_1}(\mathbf{k}, \mathbf{k}_1) q_0^{s_2 s_3}(\mathbf{k}_2, \mathbf{k}_3) \delta^{0 1}(\mathbf{k}) \delta^{2 3}(\mathbf{k}) \Delta(\Omega^{0'}_{123}, T_0) \delta^{0'}_{123}(\mathbf{k}) \prod_{i=1}^3 \mathrm{d}^n\mathbf{k}_i \\
    & \hspace{-40pt} + \int_{\mathbb{R}^{3n}} \sum_{s_1, s_2, s_3} \mathbf{L}^{-s' s_1 s_2 s_3}_{-\mathbf{k}' \mathbf{k}_1 \mathbf{k}_2 \mathbf{k}_3} q_0^{s s_1 s_2 s_3}(\mathbf{k}, \mathbf{k}_1, \mathbf{k}_2, \mathbf{k}_3) \delta^{0 1 2 3}(\mathbf{k}) \Delta(\Omega^{0'}_{123}, T_0) \delta^{0'}_{123}(\mathbf{k}) \prod_{i=1}^3 \mathrm{d}^n\mathbf{k}_i .
    \label{eq:a0b2}
\end{aligned}
\end{equation}
The first term can exhibit secular growth if we select the polarization as $s_1 = s'$ and $s_2 = -s_3$. In this case, the Dirac delta functions ensure that $\omega_{\mathbf{k}'} = \omega_{\mathbf{k}_1}$ and $\omega_{\mathbf{k}_2} = \omega_{\mathbf{k}_3}$, which leads to the vanishing of $\Omega^{0'}_{123}$. The remaining terms do not contribute to secular growth.
Therefore, we obtain the following result:
\begin{equation}
    \mathcal{C}_{T_0} \left\{ \langle a^{s}_0(\mathbf{k}) b^{s'}_2(\mathbf{k}') \rangle \right\} = 
    3 \langle a^{s}_0(\mathbf{k}) a^{s'}_0(\mathbf{k}') \rangle F(\mathbf{k}', s'),
\end{equation}
where:
\begin{equation}
    F(\mathbf{k}', s') = \int_{\mathbb{R}^n} \sum_{s_2} \mathbf{L}^{-s' s' s_2 -s_2}_{-\mathbf{k}' \mathbf{k}' \mathbf{k}_2 -\mathbf{k}_2} q_0^{s_2 -s_2}(\mathbf{k}_2, -\mathbf{k}_2)~\mathrm{d}^n\mathbf{k}_2.
\end{equation}

Combining both secular contributions, we have:
\begin{equation}
	\partial_{T_2} \langle a^{s}_0(\mathbf{k}) a^{s'}_0(\mathbf{k}') \rangle =\\
    3 \langle a^{s}_0(\mathbf{k}) a^{s'}_0(\mathbf{k}') \rangle \left[ F(\mathbf{k}, s) + F(\mathbf{k}', s')\right].
\end{equation}
However, recalling that we are in the case where $\mathbf{k}' = - \mathbf{k}$ and $s' = - s$ which is the only relevant case for our study, we may notice that:
\begin{equation}
\begin{aligned}
    F(\mathbf{k}', s') &= \int_{\mathbb{R}^n} \sum_{s_2} \mathbf{L}^{s -s s_2 -s_2}_{\mathbf{k} -\mathbf{k} \mathbf{k}_2 -\mathbf{k}_2} q_0^{s_2 -s_2}(\mathbf{k}_2, -\mathbf{k}_2)~\mathrm{d}^n\mathbf{k}_2 \\
    &= - \int_{\mathbb{R}^n} \sum_{s_2} \mathbf{L}^{-s s s_2 -s_2}_{-\mathbf{k} \mathbf{k} \mathbf{k}_2 -\mathbf{k}_2} q_0^{s_2 -s_2}(\mathbf{k}_2, -\mathbf{k}_2)~\mathrm{d}^n\mathbf{k}_2 \\
    &= - F(\mathbf{k}, s)
\end{aligned}
\end{equation}
which implies that $\partial_{T_2} \langle a^{s}_0(\mathbf{k}) a^{s'}_0(\mathbf{k}') \rangle$ vanishes when we consider the symmetries of the interaction, resulting in:
\begin{equation}
	\partial_{T_2} \langle a^{s}_0(\mathbf{k}) a^{s'}_0(\mathbf{k}') \rangle = 0.
	\label{eq:dT2wa}
\end{equation}
Once again, the wave action does not evolve with respect to $T_2$ and the turbulent cascade does not develop at this short timescale. 

This analysis can be extended to the fourth-order moment. We obtain:
\begin{multline}
    \partial_{T_2} \langle a^{s}_0(\mathbf{k}) a^{s'}_0(\mathbf{k}') a^{s''}_0(\mathbf{k}'') a^{s'''}_0(\mathbf{k}''') \rangle \\
    = 3\langle a^{s}_0(\mathbf{k}) a^{s'}_0(\mathbf{k}') a^{s''}_0(\mathbf{k}'') a^{s'''}_0(\mathbf{k}''') \rangle \left[ F(\mathbf{k}, s) + F(\mathbf{k}', s') + F(\mathbf{k}'', s'') + F(\mathbf{k}''', s''') \right].
\end{multline}
From the second closure, we know that the second-order moment does not depend\footnote{This result has been proved when $s' = -s$. In the other case, where $s' = s$ the statistical mean $\langle a^s(\mathbf{k}, t) a^{s'}(\mathbf{k}', t) e^{i (s \omega_\mathbf{k} + s' \omega_{\mathbf{k}'}) t}\rangle = q(\mathbf{k}, \mathbf{k}') \delta(\mathbf{k} + \mathbf{k}') \langle e^{i (s \omega_\mathbf{k} + s' \omega_{\mathbf{k}'}) t} \rangle$ is necessary zero since the dirac function impose that $\omega_\mathbf{k} = \omega_{\mathbf{k}'}$ so oscillating term vanishes. Thus, the case $s'=s$ has no relevant physical meaning.} on $T_2$, and therefore the left hand side can be reduced to a fourth-order cumulant. On the right hand side, we see that the decomposition in terms of second-order moments does not give any contribution because of the pairwise cancellation. Therefore, the previous equation can be reduced to:
\begin{multline}
    \partial_{T_2} 
    q_0^{s s' s'' s'''}(\mathbf{k}, \mathbf{k'}, \mathbf{k''}, \mathbf{k'''}) \\
    = 3q_0^{s s' s'' s'''}(\mathbf{k}, \mathbf{k'}, \mathbf{k''}, \mathbf{k'''}) \left[ F(\mathbf{k}, s) + F(\mathbf{k}', s') + F(\mathbf{k}'', s'') + F(\mathbf{k}''', s''') \right].
\end{multline}
This relation can be integrated according to $T_2$ timescale, giving:
\begin{multline}
    q_0^{s s' s'' s'''}(\mathbf{k}, \mathbf{k'}, \mathbf{k''}, \mathbf{k'''}) \\
    = \left. q_0^{s s' s'' s'''}(\mathbf{k}, \mathbf{k'}, \mathbf{k''}, \mathbf{k'''}) \right|_{t=0} \exp \left\{ 3 T_2 \left[ F(\mathbf{k}, s) + F(\mathbf{k}', s') + F(\mathbf{k}'', s'') + F(\mathbf{k}''', s''') \right] \right\}.
\end{multline}

However, according to the initial conditions (Gaussian statistics or random phase assumption)
\begin{equation}
    \left. q_0^{s s' s'' s'''}(\mathbf{k}, \mathbf{k'}, \mathbf{k''}, \mathbf{k'''}) \right|_{t=0} = 0,
\end{equation}
which implies that:
\begin{equation}
 q_0^{s s' s'' s'''}(\mathbf{k}, \mathbf{k'}, \mathbf{k''}, \mathbf{k'''}) = 0,
    \label{eq:dT2wa2}
\end{equation}
and its first derivative in $T_2$ is also equal to zero.
Therefore, we have proved that this Gaussian/random phase property assumed initially propagates (at least) up to $T_2$ when initially assumed.

\subsection{Third asymptotic closure at time $\mathbf{T_3}$}

At next order, we have:
\begin{equation}
\begin{aligned}
	\langle a^{s}_0(\mathbf{k}) a^{s'}_3(\mathbf{k}') +  a^{s}_1(\mathbf{k}) a^{s'}_2(\mathbf{k}') + 
    a^{s}_2(\mathbf{k}) a^{s'}_1(\mathbf{k}') + a^{s}_3(\mathbf{k}) a^{s'}_0(\mathbf{k}')\rangle & \\
    & \hspace{-250pt}= - T_0 \partial_{T_3} \langle a^{s}_0(\mathbf{k}) a^{s'}_0(\mathbf{k}') \rangle + T_0^{\phantom{a}2} \partial_{T_1} \partial_{T_2} \langle a^{s}_0(\mathbf{k}) a^{s'}_0(\mathbf{k}') \rangle 
	- \frac{T_0^{\phantom{a}3}}{6} \partial_{T_1}^3 \langle a^{s}_0(\mathbf{k}) a^{s'}_0(\mathbf{k}') \rangle \\
	& \hspace{-238pt} - T_0 \partial_{T_1} \langle a^{s}_0(\mathbf{k}) b^{s'}_2(\mathbf{k}') + a^{s'}_0(\mathbf{k}') b^{s}_2(\mathbf{k}) \rangle.
\end{aligned}
\end{equation}
From the first closure (\ref{eq:dT1wa}), the contributions proportional to $T_0^2$ and $T_0^3$ vanish. The right-hand side of the equation involves terms of the form:
\begin{equation}
	\partial_{T_1} \langle a^{s}_0(\mathbf{k}) a^{s'}_0(\mathbf{k}') a^{s''}_0(\mathbf{k}'') a^{s'''}_0(\mathbf{k}''')\rangle,
\end{equation}
which vanishes according to eq. (\ref{eq:dT1wa2}). Therefore, in order to maintain a bounded left-hand side, we must impose the asymptotic condition:
\begin{equation}
	\partial_{T_3} \langle a^{s}_0(\mathbf{k}) a^{s'}_0(\mathbf{k}') \rangle = 0,
	\label{eq:dT3wa}
\end{equation}
hence, the wave action does not evolve with respect to $T_3$ either. In other words, this timescale is still too short to allow the development of a turbulent cascade. As we will see in the next section, wave turbulence develops on a timescale $T_4$.

\subsection{Fourth asymptotic closure at time $\mathbf{T_4}$}

At order $\epsilon^4$, we have the following expression:
\begin{equation}
\begin{aligned}
	\langle a^{s}_0(\mathbf{k}) a^{s'}_4(\mathbf{k}') + a^{s}_1(\mathbf{k}) a^{s'}_3(\mathbf{k}') + a^{s}_2(\mathbf{k}) a^{s'}_2(\mathbf{k}') + a^{s}_3(\mathbf{k}) a^{s'}_1(\mathbf{k}') +
	a^{s}_4(\mathbf{k}) a^{s'}_0(\mathbf{k}') \rangle & \\
    & \hspace{-350pt} = - T_0 \partial_{T_4} \langle a^{s}_0(\mathbf{k}) a^{s'}_0(\mathbf{k}') \rangle 
	+ \frac{T_0^{\phantom{a}2}}{2} \left[ \partial_{T_1}\partial_{T_3} \langle a^{s}_0(\mathbf{k}) a^{s'}_0(\mathbf{k}') \rangle + \partial_{T_2}^2 \langle a^{s}_0(\mathbf{k}) a^{s'}_0(\mathbf{k}') \rangle \right] \\
    & \hspace{-340pt} - \frac{T_0^{\phantom{a}3}}{2} \partial_{T_1}^2 \partial_{T_2} \langle a^{s}_0(\mathbf{k}) a^{s'}_0(\mathbf{k}')\rangle + \frac{T_0^{\phantom{a}4}}{24} \partial_{T_1}^4 \langle  a^{s}_0(\mathbf{k}) a^{s'}_0(\mathbf{k}') \rangle \\
    & \hspace{-340pt} + \langle b^{s}_2(\mathbf{k}) b^{s'}_2(\mathbf{k}')\rangle + \langle a^{s}_0(\mathbf{k}) b^{s'}_4(\mathbf{k}') + a^{s'}_0(\mathbf{k}') b^{s}_4(\mathbf{k})\rangle \\
    & \hspace{-340pt} - T_0 \partial_{T_2} \langle a_0^s(\mathbf{k}) b_2^{s'}(\mathbf{k}') + a_0^{s'}(\mathbf{k}') b_2^{s}(\mathbf{k}) \rangle 
	+ \frac{T_0^{\phantom{a}2}}{2} \partial_{T_1}^2 \langle a^{s}_0(\mathbf{k}) b^{s'}_2(\mathbf{k}') + a^{s'}_0(\mathbf{k}') b^{s}_2(\mathbf{k}) \rangle.
    \label{eq:KEp}
\end{aligned}
\end{equation}
Using eq. (\ref{eq:dT1wa}), (\ref{eq:dT1wa2}), (\ref{eq:dT2wa}), and (\ref{eq:dT2wa2}), we can simplify the previous relation significantly and obtain:
\begin{multline}
	\langle a^{s}_0(\mathbf{k}) a^{s'}_4(\mathbf{k}') + a^{s}_1(\mathbf{k}) a^{s'}_3(\mathbf{k}') + a^{s}_2(\mathbf{k}) a^{s'}_2(\mathbf{k}') + a^{s}_3(\mathbf{k}) a^{s'}_1(\mathbf{k}') + a^{s}_4(\mathbf{k}) a^{s'}_0(\mathbf{k}') \rangle \\
	\hspace{-300pt}= - T_0 \partial_{T_4} \langle a^{s}_0(\mathbf{k}) a^{s'}_0(\mathbf{k}') \rangle + \langle b^{s}_2(\mathbf{k}) b^{s'}_2(\mathbf{k}')\rangle + \langle a^{s}_0(\mathbf{k}) b^{s'}_4(\mathbf{k}') + a^{s'}_0(\mathbf{k}') b^{s}_4(\mathbf{k})\rangle .
\end{multline}
Interestingly, in the expression (\ref{eq:KEp}), the cancellation of the term involving the $T_2$ time derivative is due to the Gaussian assumption made at $t=0$.

According to the Riemann-Lebesgue lemma, both terms $\langle a^{s}_0(\mathbf{k}) b^{s'}_4(\mathbf{k}') + b^{s}_4(\mathbf{k}) a^{s'}_0(\mathbf{k}') \rangle$ and $\langle b^{s}_2(\mathbf{k}) b^{s'}_2(\mathbf{k}') \rangle$ exhibit secular divergences. Thus, the evolution of the wave action is given by:
\begin{equation}
	\partial_{T_4} \langle a^{s}_0(\mathbf{k}) a^{s'}_0(\mathbf{k}') \rangle = \mathcal{C}_{T_0} \left\{ \langle b^{s}_2(\mathbf{k}) b^{s'}_2(\mathbf{k}')\rangle + \langle a^{s}_0(\mathbf{k}) b^{s'}_4(\mathbf{k}') + a^{s'}_0(\mathbf{k}') b^{s}_4(\mathbf{k}) \rangle \right\}.
\label{eq:dT4wa}
\end{equation}

The subtlety here is that the resonance can be twofold. In the uni-resonant case, these terms show a linear secular drift in $T_0$, while in the bi-resonant case, they exhibit a drift in $T_0^2$. But it will be shown later that:
\begin{equation}
	\mathcal{C}_{T_0^{2}} \left\{ \langle b^{s}_2(\mathbf{k}) b^{s'}_2(\mathbf{k}')\rangle + \langle a^{s}_0(\mathbf{k}) b^{s'}_4(\mathbf{k}') + a^{s'}_0(\mathbf{k}') b^{s}_4(\mathbf{k}) \rangle \right\} = 0.
	\label{eq:compatibilite_energie}
\end{equation}

The sixth-order moment $\langle a^{s_1}_0(\mathbf{k}_1) a^{s_2}_0(\mathbf{k}_2) a^{s_3}_0(\mathbf{k}_3) a^{s_4}_0(\mathbf{k}_4) a^{s_5}_0(\mathbf{k}_5) a^{s_6}_0(\mathbf{k}_6) \rangle$ can be written as a sum of different cumulants of the form (see Appendix \ref{sec:appendix_six_points_cummulant}):
\begin{itemize}
    \item $q_0^{s_1 s_2}(\mathbf{k}_1, \mathbf{k}_2) q_0^{s_3 s_4}(\mathbf{k}_3, \mathbf{k}_4) q_0^{s_5 s_6}(\mathbf{k}_5, \mathbf{k}_6) \delta^{12}(\mathbf{k}) \delta^{34}(\mathbf{k}) \delta^{56}(\mathbf{k})$ and other permutations;
    \item $q_0^{s_1 s_2}(\mathbf{k}_1, \mathbf{k}_2) q_0^{s_3 s_4 s_5 s_6}(\mathbf{k}_3, \mathbf{k}_4, \mathbf{k}_5, \mathbf{k}_6) \delta^{12}(\mathbf{k}) \delta^{3456}(\mathbf{k})$ and other permutations;
    \item $q_0^{s_1 s_2 s_3}(\mathbf{k}_1, \mathbf{k}_2, \mathbf{k}_3) q_0^{s_4 s_5 s_6}(\mathbf{k}_4, \mathbf{k}_5, \mathbf{k}_6) \delta^{123}(\mathbf{k}) \delta^{456}(\mathbf{k})$ and other permutations;
    \item $q_0^{s_1 s_2 s_3 s_4 s_5 s_6}(\mathbf{k}_1, \mathbf{k}_2, \mathbf{k}_3, \mathbf{k}_4, \mathbf{k}_5, \mathbf{k}_6) \delta^{123456}(\mathbf{k})$.
\end{itemize}
Therefore, we need to examine all the different contributions (41 terms) to determine which ones could lead to a secular divergence. We can immediately disregard the contribution of $q_0^{s_1 s_2 s_3 s_4 s_5 s_6}(\mathbf{k}_1, \mathbf{k}_2, \mathbf{k}_3, \mathbf{k}_4, \mathbf{k}_5, \mathbf{k}_6)$ $\delta^{123456}(\mathbf{k})$ and terms of the form $q_0^{s_1 s_2 s_3}(\mathbf{k}_1, \mathbf{k}_2, \mathbf{k}_3) \delta^{123}(\mathbf{k})$ $q_0^{s_4 s_5 s_6}(\mathbf{k}_4, \mathbf{k}_5, \mathbf{k}_6) \delta^{456}(\mathbf{k})$. This is because the different Dirac deltas do not impose any constraints on the integration over the resonant manifold.

\subsubsection*{Study of the term $\langle a^{s}_0(\mathbf{k}) b^{s'}_4(\mathbf{k}') + b^{s}_4(\mathbf{k}) a^{s'}_0(\mathbf{k}')\rangle$}

Using the theory of generalized functions, different types of oscillating integrals can be evaluated, which generalizes the Riemann-Lebesgue lemma. In particular, 
we can find the following long time limits of $E(X, Y, T)$ (see Appendix \ref{sec:consequences_lemma}):
\begin{equation}
E(X,Y,T) \xrightarrow[T \to + \infty]{} 
\begin{cases}
	\left[\pi \delta(X) + i \mathcal{P} \left( \frac{1}{X} \right)\right] \left[\pi \delta(Y) + i \mathcal{P} \left( \frac{1}{Y} \right)\right], & \text{if}~X, Y \neq 0,~Y\text{and}~X\neq Y;\\
	\left[\pi \delta(X) + i \mathcal{P} \left( \frac{1}{X} \right)\right] \left[T - i \frac{\partial}{\partial X}\right], & \text{if}~X \neq 0~\text{and}~Y = 0;\\
	\left[\pi \delta(Y) + i \mathcal{P} \left( \frac{1}{Y} \right)\right] \left[T - i \frac{\partial}{\partial Y}\right], & \text{if}~X = 0~\text{and}~Y \neq 0;\\
	\left[\pi \delta(X) + i \mathcal{P} \left( \frac{1}{X} \right)\right] i \frac{\partial}{\partial X},  & \text{if}~X = Y \neq 0;\\
	\frac{T^2}{2}, &\text{if}~X = Y = 0  .
\end{cases}
\label{eq:lemme_RiemannLebesgue_gal}
\end{equation}
We find that $\langle a^{s}_0(\mathbf{k}) b^{s'}_4(\mathbf{k}') \rangle$ exhibits a secular divergence if the integration is constrained to one of the following manifolds:
\begin{subequations}
\begin{equation}
\left\{
\begin{aligned}
    s' \omega' &= s_1 \omega_1 + s_2 \omega_2 + s_3 \omega_3 \\
    s' \omega' &\neq s_2 \omega_2 + s_3 \omega_3 + s_4 \omega_4 + s_5 \omega_5 + s_6 \omega_6;
\end{aligned}
\right.
\label{eq:resonnance_case1}  
\end{equation}
or,
\begin{equation}
\left\{
\begin{aligned}
    s' \omega' &\neq s_1 \omega_1 + s_2 \omega_2 + s_3 \omega_3 \\
    s' \omega' &= s_2 \omega_2 + s_3 \omega_3 + s_4 \omega_4 + s_5 \omega_5 + s_6 \omega_6. 
\end{aligned}
\right.
\label{eq:resonnance_case2}
\end{equation}
\end{subequations}
There are two types of terms that satisfy the conditions (\ref{eq:resonnance_case1}). The first type is proportional to $q_0^{s_2 s_3}(\mathbf{k}_2, \mathbf{k}_3) \delta^{23}(\mathbf{k})$, and the second type is proportional to $q_0^{s s_2}(\mathbf{k}, \mathbf{k}_2) \delta^{02}(\mathbf{k})$ (or $q_0^{s s_3}(\mathbf{k}, \mathbf{k}3) \delta^{03}(\mathbf{k})$). It is important to note that such a decomposition must satisfy $\Omega^1_{4 5 6} \neq 0$. The contribution from the first type of term is given by:
\begin{multline}
    3 \delta^{0 0'}(\mathbf{k}) F(\mathbf{k}', s') \int_{\mathbb{R}^{3n}} \sum_{\substack{s_4, s_5, s_6\\\Omega^{0'}_{4 5 6} \neq 0}} \mathbf{L}^{-s' s_4 s_5 s_6}_{-\mathbf{k}' \mathbf{k}_4 \mathbf{k}_5 \mathbf{k}_6}
    \langle a^{s}_0(\mathbf{k}) a^{s_4}_0(\mathbf{k}_4) a^{s_5}_0(\mathbf{k}_5) a^{s_6}_0(\mathbf{k}_6) \rangle \\[-20pt]
    \times \delta^{0'}_{456}(\mathbf{k}) \Delta(\Omega^{0'}_{456},T_0)~\prod_{i=4}^6\mathrm{d}^n\mathbf{k}_i .
\end{multline}
On the other hand, the contribution from the second type of term (symmetrically) is given by:
\begin{multline}
    6 \langle a_0^s(\mathbf{k}) a_0^{s'}(\mathbf{k}') \rangle 
    \int_{\mathbb{R}^{4n}}\sum_{\substack{s_1, s_4, s_5, s_6\\\Omega^{1}_{4 5 6} \neq 0}} \mathbf{L}^{-s' s' s_1 -s_1}_{-\mathbf{k}' \mathbf{k}' \mathbf{k}_1 -\mathbf{k}_1} \mathbf{L}^{-s_1 s_4 s_5 s_6}_{-\mathbf{k}_1 \mathbf{k}_4 \mathbf{k}_5 \mathbf{k}_6} \langle a^{s_1}_0(\mathbf{k}_1) a^{s_4}_0(\mathbf{k}_4) a^{s_5}_0(\mathbf{k}_5) a^{s_6}_0(\mathbf{k}_6)\rangle \\[-20pt]
    \times \delta^{1}_{456}(\mathbf{k}) \Delta(\Omega^{1}_{456},T_0)~\mathrm{d}^n\mathbf{k}_1 \prod_{i=4}^6\mathrm{d}^n\mathbf{k}_i.
\end{multline}

A decomposition that satisfies the conditions (\ref{eq:resonnance_case2}) can be expressed as:
\begin{equation}
	\langle a^{s}_0(\mathbf{k}) a^{s_i}_0(\mathbf{k}_i) \rangle \langle a^{s_2}_0(\mathbf{k}_2) a^{s_j}_0(\mathbf{k}_j) \rangle \langle a^{s_3}_0(\mathbf{k}_3) a^{s_k}_0(\mathbf{k}_k) \rangle,
\end{equation}
where $i, j$, and $k$ are distinct elements chosen from ${4, 5, 6}$. It is important to note that due to symmetry, these terms will yield the same result, as there are six of them (corresponding to the permutations of a three-element set). The total contribution from this term is:
\begin{multline}
	18  q_0^{s s'}(\mathbf{k},\mathbf{k}') \delta^{0 0'}(\mathbf{k}) \delta^{ss'} \int_{\mathbb{R}^{3n}}\sum_{\substack{s_1, s_2, s_3 \\ \Omega^{0'}_{123} \neq 0}} \mathbf{L}^{-s' s_1 s_2 s_3}_{-\mathbf{k}' \mathbf{k}_1 \mathbf{k}_2 \mathbf{k}_3} \mathbf{L}^{-s_1 s' -s_2 -s_3}_{-\mathbf{k}_1 \mathbf{k}' -\mathbf{k}_2 -\mathbf{k}_3} q_0^{s_2 -s_2}(\mathbf{k}_2, -\mathbf{k}_2) q_0^{s_3 -s_3}(\mathbf{k}_3, -\mathbf{k}_3) \\[-20pt]
    \times \delta^{0'}_{123}(\mathbf{k}) \Delta (\Omega^{0'}_{123})~ \prod_{i=1}^3\mathrm{d}^n\mathbf{k}_i.
\end{multline}

In the case of a bi-resonance, $\langle a^{s}_0(\mathbf{k}) b^{s'}_4(\mathbf{k}')\rangle$ exhibits a quadratic divergence in $T_0^2$, which occurs when the following conditions are satisfied:
\begin{equation}
\left\{
\begin{aligned}
    s' \omega' &= s_1 \omega_1 + s_2 \omega_2 + s_3 \omega_3 \\
    s_1 \omega_1 &= s_4 \omega_4 + s_5 \omega_5 + s_6 \omega_6
\end{aligned}
 \right. .
\end{equation}
This can only be achieved through a decomposition of the form:
\begin{equation*}
	\langle a^{s}_0(\mathbf{k}) a^{s_i}_0(\mathbf{k}_i) \rangle \langle a^{s_2}_0(\mathbf{k}_2) a^{s_3}_0(\mathbf{k}_3) \rangle \langle a^{s_j}_0(\mathbf{k}_j) a^{s_k}_0(\mathbf{k}_k) \rangle,
\end{equation*}
where $i, j$, and $k$ are distinct elements chosen from ${4, 5, 6}$ in pairs. The decomposition yields only three terms, which, due to symmetries, give the same result. Furthermore, the quadratic contribution of $\langle a^{s}_0(\mathbf{k}) b^{s'}_4(\mathbf{k}')\rangle$ is given by:
\begin{equation}
	\frac{9}{2} \langle a_0^s(\mathbf{k}) a_0^{s'}(\mathbf{k}') \rangle F(\mathbf{k}', s')^2.
\end{equation}

\subsubsection*{Study of the term $\langle b^{s'}_2(\mathbf{k}') b^{s}_2(\mathbf{k})\rangle$}

We will employ Riemann-Lebesgue's lemma and the following corollary (see Appendix \ref{sec:consequences_lemma}):
\begin{equation}
	\Delta(X,T) \Delta(-X,T) \xrightarrow[T \to + \infty]{} 2 \pi T \delta(X) + 2 \mathcal{P}\left(\frac{1}{X}\right) \frac{\partial}{\partial X}.
\end{equation}
The term $\langle b^s_2(\mathbf{k}) b^{s'}_2(\mathbf{k}')\rangle$ can exhibit a secular drift proportional to $T_0$ if any of the following conditions are satisfied:
\begin{subequations}
\begin{equation}
\left\{
\begin{aligned}
    s \omega &= s_1 \omega_1 + s_2 \omega_2 + s_3 \omega_3 \\
    s' \omega' &\neq s_4 \omega_4 + s_5 \omega_5 + s_6 \omega_6
\end{aligned}
\right. ;
\label{eq:resonnance_cas3}
\end{equation}
or,
\begin{equation}
\left\{
\begin{aligned}
    s \omega &\neq s_1 \omega_1 + s_2 \omega_2 + s_3 \omega_3 \\
    s' \omega' &= s_4 \omega_4 + s_5 \omega_5 + s_6 \omega_6
\end{aligned}
\right. ;
\label{eq:resonnance_cas4}
\end{equation}
or,
\begin{equation}
    s\omega - s_1 \omega_1 - s_2 \omega_2 - s_3 \omega_3 = - s' \omega' + s_4 \omega_4 + s_5 \omega_5 + s_6 \omega_6.
\label{eq:resonnance_cas5}
\end{equation}
\end{subequations}

The situation (\ref{eq:resonnance_cas3}) arises for decomposition of the form:
\begin{equation}
	\langle a^{s_i}_0(\mathbf{k}_i) a^{s_j}_0(\mathbf{k}_j)\rangle \langle a^{s_k}_0(\mathbf{k}_k) a^{s_4}_0(\mathbf{k}_4) a^{s_5}_0(\mathbf{k}_5) a^{s_6}_0(\mathbf{k}_6) \rangle,
\end{equation}
where $i$, $j$, and $k$ are elements of $\left\{ 1, 2, 3 \right\}$ pairwise distinct. These three decomposition yield the same result. Therefore, their contribution is:
\begin{multline}
3 \delta^{0 0'}(\mathbf{k}) F(\mathbf{k}, s) \int_{\mathbb{R}^{3n}} \sum_{\substack{s_4, s_5, s_6\\\Omega^{0'}_{4 5 6} \neq 0}} \mathbf{L}^{-s' s_4 s_5 s_6}_{-\mathbf{k}' \mathbf{k}_4 \mathbf{k}_5 \mathbf{k}_6} \langle a^{s}_0(\mathbf{k}) a^{s_4}_0(\mathbf{k}_4) a^{s_5}_0(\mathbf{k}_5) a^{s_6}_0(\mathbf{k}_6) \rangle \\[-20pt]
\times \delta^{0'}_{456}(\mathbf{k}) \Delta(\Omega^{0'}_{456}, T_0) ~\prod_{i=4}^6\mathrm{d}^n\mathbf{k}_i.
\end{multline}

In the same way, we can determine the contribution corresponding to condition (\ref{eq:resonnance_cas4}). This one is equal to:
\begin{multline}
3 \delta^{0 0'}(\mathbf{k}) F(\mathbf{k}', s') \int_{\mathbb{R}^{3n}} \sum_{\substack{s_1, s_2, s_3\\\Omega^{0}_{123} \neq 0}} \mathbf{L}^{-s s_1 s_2 s_3}_{-\mathbf{k} \mathbf{k}_1 \mathbf{k}_2 \mathbf{k}_3} \langle a^{s'}_0(\mathbf{k}') a^{s_1}_0(\mathbf{k}_1) a^{s_2}_0(\mathbf{k}_2) a^{s_3}_0(\mathbf{k}_3) \rangle \\[-20pt]
\times \delta^{0}_{123}(\mathbf{k}) \Delta(\Omega^{0}_{123}, T_0) ~\prod_{i=1}^3\mathrm{d}^n\mathbf{k}_i.
\end{multline}

The situation described by eq. (\ref{eq:resonnance_cas5}) requires a more careful treatment. It can only occur in a decomposition of the form:
\begin{equation}
	\langle a^{s_{i_1}}_0(\mathbf{k}_{i_1}) a^{s_{i_2}}_0(\mathbf{k}_{i_2})\rangle \langle a^{s_{j_1}}_0(\mathbf{k}_{j_1}) a^{s_{j_2}}_0(\mathbf{k}_{j_2}) \rangle \langle a^{s_{k_1}}_0(\mathbf{k}_{k_1}) a^{s_{k_2}}_0(\mathbf{k}_{k_2}) \rangle,
\end{equation}
where $i_1$, $j_1$, and $k_1$ (respectively $i_2$, $j_2$, and $k_2$) are elements of $\left\{1, 2, 3\right\}$ (respectively $\left\{4, 5, 6\right\}$) pairwise distinct. Out of the 15 terms, only 6 satisfy this condition, and due to symmetry, they yield the same result. Consequently, the total contribution is:
\begin{multline}
	12 \pi \delta^{0 0'}(\mathbf{k}) \delta^{s s'} \int_{\mathbb{R}^{3n}} \sum_{s_1, s_2, s_3} \mathbf{L}^{-s s_1 s_2 s_3}_{-\mathbf{k} \mathbf{k}_1 \mathbf{k}_2 \mathbf{k}_3} \mathbf{L}^{-s' -s_1 -s_2 -s_3}_{-\mathbf{k}' -\mathbf{k}_1 -\mathbf{k}_2 -\mathbf{k}_3} q_0^{s_1 - s_1}(\mathbf{k}_1, -\mathbf{k}_1) q_0^{s_2 - s_2}(\mathbf{k}_2, -\mathbf{k}_2) \\[-20pt]
    \times q_0^{s_3 - s_3}(\mathbf{k}_3, -\mathbf{k}_3) \delta(\Omega^{0}_{123}) \delta^0_{123}(\mathbf{k})~\prod_{i=1}^3\mathrm{d}^n\mathbf{k}_i.
\end{multline}

Finally, the term $\langle b^s_2(\mathbf{k}) b^{s'}2(\mathbf{k}')\rangle$ can exhibit a quadratic secular drift in the case of the bi-resonance, where:
\begin{equation}
\left\{
\begin{aligned}
    s \omega &= s_1 \omega_1 + s_2 \omega_2 + s_3 \omega_3 \\
    s' \omega' &= s_4 \omega_4 + s_5 \omega_5 + s_6 \omega_6
\end{aligned}
 \right. .
\end{equation}
This situation arises only for decomposition of the form:
\begin{equation}
	\langle a^{s_{i_1}}_0(\mathbf{k}_{i_1}) a^{s_{i_2}}_0(\mathbf{k}_{i_2})\rangle \langle a^{s_{j_1}}_0(\mathbf{k}_{j_1}) a^{s_{k_1}}_0(\mathbf{k}_{k_1}) \rangle \langle a^{s_{j_2}}_0(\mathbf{k}_{j_2}) a^{s_{k_2}}_0(\mathbf{k}_{k_2}) \rangle,
\end{equation}
where $i_1$, $j_1$, and $k_1$ (respectively, $i_2$, $j_2$, and $k_2$) are elements of $\left\{1, 2, 3\right\}$ (respectively, $\left\{4, 5, 6\right\}$) pairwise distinct. There are nine such decompositions that yield the same result due to symmetry. Hence, the total contribution of this term is given by:
\begin{equation}
	9 \langle a_0^s(\mathbf{k}) a_0^s(\mathbf{k}') \rangle F(\mathbf{k}, s) F(\mathbf{k}', s').
\end{equation}

\subsection{Kinetic equation at time $\mathbf{T_4}$}

All the previous results can be summarized and simplified using the symmetry properties of the interaction term. In particular, we have:
\begin{equation}
	\mathbf{L}^{s' s' s_2 -s_2}_{\mathbf{k}' \mathbf{k}' \mathbf{k}_2 -\mathbf{k}_2} = - \mathbf{L}^{s s s_2 -s_2}_{\mathbf{k} \mathbf{k} \mathbf{k}_2 -\mathbf{k}_2}.
\end{equation}
Hence, we finally find the following expression for the secular contributions:
\begin{equation}
\begin{aligned}
	\mathcal{C}_{T_0} \left\{ \langle b^{s}_2(\mathbf{k}) b^{s'}_2(\mathbf{k}')\rangle + \langle a^{s}_0(\mathbf{k}) b^{s'}_4(\mathbf{k}') + a^{s'}_0(\mathbf{k}') b^{s}_4(\mathbf{k}) \rangle \right\} & \\
	& \hspace{-225pt} = 12\pi~\delta^{0 0'}(\mathbf{k}) \int_{\mathbb{R}^{3n}} \sum_{s_1, s_2, s_3} \mathbf{L}^{-s s_1 s_2 s_3}_{-\mathbf{k} \mathbf{k}_1 \mathbf{k}_2 \mathbf{k}_3} \mathbf{L}^{-s' -s_1 -s_2 -s_3}_{-\mathbf{k}' -\mathbf{k}_1 -\mathbf{k}_2 -\mathbf{k}_3} q_0^{s_1 - s_1}(\mathbf{k}_1, -\mathbf{k}_1) \\[-15pt]
    & \hspace{-90pt} \times q_0^{s_2 - s_2}(\mathbf{k}_2, -\mathbf{k}_2) q_0^{s_3 - s_3}(\mathbf{k}_3, -\mathbf{k}_3) \delta(\Omega^{0}_{123}) \delta^0_{123}(\mathbf{k})~\prod_{i=1}^3\mathrm{d}^n\mathbf{k}_i \\
    & \hspace{-215pt} + 18 q_0^{s s'}(\mathbf{k}, \mathbf{k}') \delta^{00'}(\mathbf{k}) \int_{\mathbb{R}^{3n}} \sum_{s_1, s_2, s_3} \mathbf{L}^{-s s_1 s_2 s_3}_{-\mathbf{k} \mathbf{k}_1 \mathbf{k}_2 \mathbf{k}_3} \mathbf{L}^{-s_1 s -s_2 -s_3}_{-\mathbf{k}_1 \mathbf{k} -\mathbf{k}_2 -\mathbf{k}_3} q_0^{s_2 -s_2}(\mathbf{k}_2, -\mathbf{k}_2) \\[-15pt]
    & \hspace{-40pt} \times  q_0^{s_3 -s_3}(\mathbf{k}_3, -\mathbf{k}_3) \delta^{0}_{123}(\mathbf{k}) \Delta(\Omega^{0}_{123},T_0)~ \prod_{i=1}^3\mathrm{d}^n\mathbf{k}_i \\
	& \hspace{-215pt} + 18~q_0^{s s'}(\mathbf{k}, \mathbf{k}') \delta^{00'}(\mathbf{k}) \int_{\mathbb{R}^{3n}}\sum_{s_1, s_2, s_3} \mathbf{L}^{-s' s_1 s_2 s_3}_{-\mathbf{k}' \mathbf{k}_1 \mathbf{k}_2 \mathbf{k}_3} \mathbf{L}^{-s_1 s' -s_2 -s_3}_{-\mathbf{k}_1 \mathbf{k}' -\mathbf{k}_2 -\mathbf{k}_3} q_0^{s_2 -s_2}(\mathbf{k}_2, -\mathbf{k}_2) \\[-15pt] 
    & \hspace{-40pt} \times q_0^{s_3 -s_3}(\mathbf{k}_3, -\mathbf{k}_3) \delta^{0'}_{123}(\mathbf{k}) \Delta(\Omega^{0'}_{123},T_0)~ \prod_{i=1}^3\mathrm{d}^n\mathbf{k}_i.
	\label{eq:contribution_lineaire_T4}
\end{aligned}
\end{equation}
Furthermore, we find that all the contributions in quadratic secular drifts in $T_0$ vanish:
\begin{equation}
	\mathcal{C}_{T_0^{\phantom{a}2}} \left\{ \langle b^{s}_2(\mathbf{k}) b^{s'}_2(\mathbf{k}')\rangle + \langle a^{s}_0(\mathbf{k}) b^{s'}_4(\mathbf{k}') + a^{s'}_0(\mathbf{k}') b^{s}_4(\mathbf{k}) \rangle \right\} = 0.
\end{equation}
So, the cumulants are well-ordered and the development is consistent. 

Before taking the limit $T_0 \to + \infty$, we introduce the wave action as: 
\begin{equation}
	n^s(\mathbf{k}) = q_0^{s-s}(\mathbf{k}, -\mathbf{k}).
\end{equation}
Next, we integrate eq. (\ref{eq:contribution_lineaire_T4}) with respect to $\mathbf{k}'$ and take the limit $T_0 \to + \infty$. Since the integrand decays sufficiently quickly (thanks to the cumulant), we can interchange the limit and the integral. This yields the following expression:
\begin{equation}
\begin{aligned}
	\partial_{T_4} &n^s(\mathbf{k}) &\\
    & = 12 \pi \int_{\mathbb{R}^{3n}} \sum_{s_1, s_2, s_3} \mathbf{L}^{-s s_1 s_2 s_3}_{-\mathbf{k} \mathbf{k}_1 \mathbf{k}_2 \mathbf{k}_3} \mathbf{L}^{s -s_1 -s_2 -s_3}_{\mathbf{k} -\mathbf{k}_1 -\mathbf{k}_2 -\mathbf{k}_3} n^{s_1}(\mathbf{k}_1) n^{s_2}(\mathbf{k}_2) n^{s_3}(\mathbf{k}_3) \delta(\Omega^{0}_{123}) \delta^0_{123}(\mathbf{k})~\prod_{i=1}^3\mathrm{d}^n\mathbf{k}_i \\[-5pt]
	& + 18 \pi n^s(\mathbf{k}) \int_{\mathbb{R}^{3n}} \sum_{s_1, s_2, s_3} \mathbf{L}^{s s_1 s_2 s_3}_{\mathbf{k} \mathbf{k}_1 \mathbf{k}_2 \mathbf{k}_3} \mathbf{L}^{-s_1 -s -s_2 -s_3}_{-\mathbf{k}_1 -\mathbf{k} -\mathbf{k}_2 -\mathbf{k}_3} n^{s_2}(\mathbf{k}_2) n^{s_3}(\mathbf{k}_3) \delta^{0123}(\mathbf{k}) \delta(\Omega^{0123})~ \prod_{i=1}^3\mathrm{d}^n\mathbf{k}_i \\[-5pt]
	& + 18 \pi n^s(\mathbf{k}) \int_{\mathbb{R}^{3n}}\sum_{s_1, s_2, s_3} \mathbf{L}^{-s s_1 s_2 s_3}_{-\mathbf{k} \mathbf{k}_1 \mathbf{k}_2 \mathbf{k}_3} \mathbf{L}^{-s_1 s -s_2 -s_3}_{-\mathbf{k}_1 \mathbf{k} -\mathbf{k}_2 -\mathbf{k}_3} n^{s_2}(\mathbf{k}_2) n^{s_3}(\mathbf{k}_3) \delta^{0}_{123}(\mathbf{k}) \delta(\Omega^{0}_{123})~ \prod_{i=1}^3\mathrm{d}^n\mathbf{k}_i \\
    & = 12 \pi \int_{\mathbb{R}^{3n}} \sum_{s_1, s_2, s_3} 
    \mathbf{L}^{s -s_1 -s_2 -s_3}_{\mathbf{k} -\mathbf{k}_1 -\mathbf{k}_2 -\mathbf{k}_3}
    \left(\frac{\mathbf{L}^{-s s_1 s_2 s_3}_{-\mathbf{k} \mathbf{k}_1 \mathbf{k}_2 \mathbf{k}_3}}{n^{s}(\mathbf{k})} + \frac{\mathbf{L}^{-s_1 s s_2 s_3}_{-\mathbf{k}_1 \mathbf{k} \mathbf{k}_2 \mathbf{k}_3}}{n^{s_1}(\mathbf{k}_1)} + \frac{\mathbf{L}^{-s_2 s s_1 s_3}_{-\mathbf{k}_2 \mathbf{k} \mathbf{k}_1 \mathbf{k}_3}}{n^{s_2}(\mathbf{k}_2)} + \frac{\mathbf{L}^{-s_3 s s_1 s_2}_{-\mathbf{k}_3 \mathbf{k} \mathbf{k}_1 \mathbf{k}_2}}{n^{s_3}(\mathbf{k}_3)} \right) \\
    & \hspace{160pt} \times n^{s}(\mathbf{k}) n^{s_1}(\mathbf{k}_1) n^{s_2}(\mathbf{k}_2) n^{s_3}(\mathbf{k}_3) \delta(\Omega^{0}_{123}) \delta^0_{123}(\mathbf{k})~\prod_{i=1}^3\mathrm{d}^n\mathbf{k}_i.
\end{aligned}
\label{eq:cinetique_general}
\end{equation}
The last equality has been computed by applying the changes of variables $\mathbf{k}'_i = - \mathbf{k}_i$ and $s'_i = - s_i$ in the second integral and by taking advantage of symmetries on the resonant manifold. This last equation represents the kinetic equation for four-wave interactions.

\section{The case of gravitational waves}
\label{sec:application}

We will apply the previous formalism to gravitational waves when the Hadad-Zakharov metric is used. Therefore, we will introduce the specific expression of the interaction term $\mathbf{L}^{s s_1 s_2 s_3}_{\mathbf{k} \mathbf{k}_1 \mathbf{k}_2 \mathbf{k}_3}$ derived earlier, into expression (\ref{eq:cinetique_general}). This will allow us to investigate the various instances of resonance in greater detail. Using the definition of normal variables in eq. (\ref{eq2p6}) it is clear that both wave action and energy do not depend on $s$. Thus, we will simply write them as $n(\mathbf{k})$ and $e(\mathbf{k}) = \omega_{\mathbf{k}} n(\mathbf{k})$, respectively. 

\subsection{Resonance conditions}

Equation (\ref{eq:cinetique_general}) provides us with the interaction conditions for a quartet of wave vectors $\left(\mathbf{k}, \mathbf{k}_1, \mathbf{k}_2, \mathbf{k}_3\right)$. These conditions are given by:
\begin{equation}
\begin{cases}
	\mathbf{k} = \mathbf{k}_1 + \mathbf{k}_2 + \mathbf{k}_3 \\
	sk = s_1 k_1 + s_2 k_2 + s_3 k_3
\end{cases}.
\end{equation}
We can identify four possible cases:
\begin{itemize}
\begin{subequations}
\item All signs are the same: 
\begin{equation}
	s = s_1 = s_2 = s_3.
	\label{eq:4+0-}
\end{equation}
\item Only one sign is different from the others, for instance:
\begin{equation}
	s = - s_1 = s_2 = s_3.
	\label{eq:3+1-}
\end{equation}
\item One sign is equal to $s$, and the other two signs are different, for instance:
\begin{equation}
	s = s_1 = - s_2 = - s_3.
	\label{eq:2+2-}
\end{equation}
\item All free signs are different from $s$:
\begin{equation}
	s = - s_1 = - s_2 = - s_3.
	\label{eq:1+3-}
\end{equation}
\end{subequations}
\end{itemize}

\subsubsection*{$4 \leftrightarrow 0$ interactions}

The $4 \leftrightarrow 0$ interactions are described by the case of eq. (\ref{eq:1+3-}). They are quickly ruled out. Indeed, in this situation the sum of the norms of the wave vectors is zero, so all the wave vectors are zero. In particular $\mathbf{k}$, which, given that $\mathbf{L}^{s s_1 s_2 s_3}_{\mathbf{0} \mathbf{k}_1 \mathbf{k}_2 \mathbf{k}_3} = 0$, implies that the contribution of this interaction vanishes.

\subsubsection*{$3 \leftrightarrow 1$ interactions}

\begin{figure}[ht]
    \centering
    \includegraphics{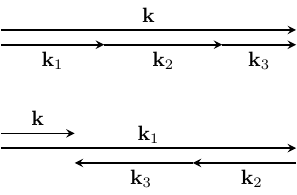}
    \caption{Diagram of the interaction mode in the case of eq. (\ref{eq:4+0-}) (on top) and in the case of eq. (\ref{eq:2+2-}) (on bottom).}
    \label{fig:3-1_interaction}
\end{figure}

The $3 \leftrightarrow 1$ interactions correspond to the cases described by eq. (\ref{eq:4+0-}) and (\ref{eq:2+2-}). In these scenarios, the wave vectors must be aligned and two situations are possible as shown in fig. (\ref{fig:3-1_interaction}). The wave vectors can then be written as follow:
\begin{equation}
    \mathbf{k} = k \begin{pmatrix} \cos \theta \\ \sin \theta \end{pmatrix} \quad \text{and} \quad \mathbf{k}_i = s_i k_i \begin{pmatrix} \cos \theta \\ \sin \theta \end{pmatrix}.
\end{equation}
This decomposition is introduced into the expression of $\mathbf{L}^{s s_1 s_2 s_3}_{\mathbf{k} \mathbf{k}_1 \mathbf{k}_2 \mathbf{k}_3}$ to show, after a direct calculation that it vanishes. By leveraging the different symmetries of the interaction coefficient, this result is also proved to be true when one sign is equal to $s$ and the other two are differents.

\subsubsection*{$2 \leftrightarrow 2$ interactions}

\begin{figure}[!ht]
    \centering
    \includegraphics{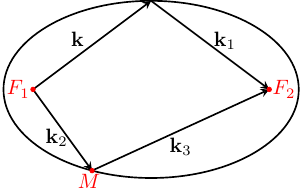}
    \caption{Schematic representation of the manifold upon $2 \leftrightarrow 2$ interactions are possible. It is simply defined by the position of a dot $M$ (in red) located on an ellipse whose foci are $\mathrm{F}_1$ and $\mathrm{F}_2$.}
    \label{fig:2-2_interaction}
\end{figure}

The $2 \leftrightarrow 2$ interactions are described by the case of the equation (\ref{eq:3+1-}). It allows much more geometries in the interaction quartet because the wave vectors are not necessarily aligned, thus the wave vectors are not simply proportional to each other. However, a change of variable in the equation (\ref{eq:cinetique_general}): $\mathbf{k}_1 \to - \mathbf{k}_1$, change the resonance condition as following:
\begin{equation}
\begin{cases}
	\mathbf{k} + \mathbf{k}_1 = \mathbf{k}_2 + \mathbf{k}_3 \\
	k + k_1 = k_2 + k_3
\end{cases}.
\label{eq:resonance_22}
\end{equation}
Therefore, for a given pair of wave vector $\left(\mathbf{k}, \mathbf{k}_1\right)$, a solution of (\ref{eq:resonance_22}) can be see as a point on an ellipse whose focus are the extremities of $\mathbf{k}$ and $\mathbf{k}_1$ (see Figure \ref{fig:2-2_interaction}).

With this representation, we note that in the limit of nonlocal interactions originating from a small wave vector (say $k \to 0$), the ellipse necessarily tends to be flat with the foci that tend to be close to the ellipse. In this limit, the three other wave vectors tend to be aligned, a situation in which the interaction coefficient vanishes. 

\subsection{Kinetic equation of gravitational waves}

Equation (\ref{eq:cinetique_general}) can be further simplified by using the following additional symmetry, valid on the resonant manifold (\ref{eq:resonance_22}) and verified numerically:
\begin{equation}
\mathbf{L}^{s s -s -s}_{\mathbf{k}_1 \mathbf{k} -\mathbf{k}_2 -\mathbf{k}3} = \mathbf{L}^{s s -s -s}_{\mathbf{k} \mathbf{k}_1 -\mathbf{k}_2 -\mathbf{k}_3}.
\end{equation}

Finally, we obtain a classical form for the kinetic equation, namely:
\begin{multline}
	\partial_t n(\mathbf{k}) =  36 \pi \epsilon^4 \int_{\mathbb{R}^6} \left| \mathbf{L}^{-s -s s s}_{-\mathbf{k} -\mathbf{k}_1 \mathbf{k}_2 \mathbf{k}_3} \right|^2 \left( \frac{1}{n(\mathbf{k})} + \frac{1}{n(\mathbf{k}_1)} - \frac{1}{n(\mathbf{k}_2)} - \frac{1}{n(\mathbf{k}_3)} \right) \\
    \times n(\mathbf{k}) n(\mathbf{k}_1) n(\mathbf{k}_2) n(\mathbf{k}_3)\delta^{01}_{23}(\omega) \delta^{01}_{23}(\mathbf{k})\prod_{i=1}^3\mathrm{d}^2\mathbf{k}_i.
	\label{eq:cinetique_particulier}
\end{multline}
We naturally use the definition $T_4 = \epsilon^4 t$, and the interaction coefficient is given by:
\begin{equation}
\begin{aligned}
    36 \pi^2 \left| \mathbf{L}^{-s -s s s}_{-\mathbf{k} -\mathbf{k}_1 \mathbf{k}_2 \mathbf{k}_3} \right|^2 &\\
    & \hspace{-75pt}
    = \frac{\pi}{4 k k_1 k_2 k_3} \Bigg\{ 
    \frac{k k_1 \left( k_3 p_2 q_2 + k_2 p_3 q_3 \right) - p p_1 \left( k_2 p_3 q_2 + k_3 p_2 q_3 \right) - q q_1 \left( k_2 p_2 q_3 + k_3 p_3 q_2 \right)}{(k_2 + k_3)(p_2 + p_3)(q_2 + q_3)} \\
    & \hspace{-12pt} 
    + \frac{k k_2 \left( k_1 p_3 q_3 - k_3 p_1 q_1 \right) + p p_2 \left( k_1 p_3 q_1 - k_3 p_1 q_3 \right) + q q_2 \left( k_1 p_1 q_3 - k_3 p_3 q_1 \right)}{(k_3 - k_1)(p_3 - p_1)(q_3 - q_1)} \\
    & \hspace{-12pt} 
    + \frac{k k_3 \left( k_1 p_2 q_2 - k_2 p_1 q_1 \right) + p p_3 \left( k_1 p_2 q_1 - k_2 p_1 q_2 \right) + q q_3 \left( k_1 p_1 q_2 - k_2 p_2 q_1 \right)}{(k_2 - k_1)(p_2 - p_1)(q_2 - q_1)} \Bigg\}^2 .
    \label{eq:coeffinter}
\end{aligned}
\end{equation}
The form of this interaction coefficient is not the same as that derived by \citep{galtier_2017}. The reason is that in the Hamiltonian derivation, an additional symmetry is introduced. However, we have verified numerically that both expressions give the same result when this additional symmetry is also introduced in expression (\ref{eq:coeffinter}). As often observed (as with capillary waves \cite{galtier_2021b}), the expression of the interaction coefficient obtained with the Eulerian derivation has a more global symmetrical appearance. In the case of GW turbulence, the difference with \cite{galtier_2017} is striking.

\subsection{Exact solutions}

Given that this problem only allows for $2 \leftrightarrow 2$ interactions, the conservation of wave action and energy becomes apparent on equation (\ref{eq:cinetique_particulier}).
Consequently, we can formulate two conservation equations in the spectral domain:
\begin{subequations}
\begin{equation}
	\partial_t n(\mathbf{k}) + \boldsymbol{\nabla}_\mathbf{k} \Xi (\mathbf{k}) = 0,
	\label{eq:conservationN}
\end{equation}
and,
\begin{equation}
	\partial_t e(\mathbf{k}) + \boldsymbol{\nabla}_\mathbf{k} \Pi (\mathbf{k}) = 0,
	\label{eq:conservationE}
\end{equation}
\end{subequations}
where $\Xi (\mathbf{k})$ and $\Pi (\mathbf{k})$ are respectively the wave action flux and the energy flux. 
Under the assumption of isotropic turbulence, we integrate angularly the previous equation, we find:
\begin{subequations}
\begin{equation}
	\partial_t N(k) + \partial_k \zeta (k) = 0,
	\label{eq:conservationN2}
\end{equation}
and,
\begin{equation}
	\partial_t E(k) + \partial_k \varepsilon (k) = 0,
	\label{eq:conservationE2}
\end{equation}
\end{subequations}
where $N(k) = \int_0^{2 \pi} k n(\mathbf{k}) \mathrm{d}\theta = 2 \pi k n(\mathbf{k})$, $E(k) = 2\pi k e(\mathbf{k})$, $\zeta (k) = 2 \pi k \Xi(\mathbf{k})$ and $\varepsilon(k) = 2 \pi k \Pi (\mathbf{k})$.
Furthermore, we use the isotropic assumption in order to determine the isotropic kinetic equation by integrating equation (\ref{eq:cinetique_particulier}) upon the angles. We have:
\begin{multline}
    \partial_t N(k) = \int_{(\mathbb{R}^+)^3} \mathcal{C}_{k k_1 k_2 k_3} \left( \frac{k}{N(k)} + \frac{k_1}{N(k_1)} - \frac{k_2}{N(k_2)} - \frac{k_3}{N(k_3)} \right) \\
    \times N(k) N(k_1) N(k_2) N(k_3) \delta^{01}_{23}(\omega) \mathrm{d}k_1 \mathrm{d}k_2 \mathrm{d}k_3,
\label{eq:conservation_N_isotrope}
\end{multline}
where:
\begin{equation}
\mathcal{C}_{k k_1 k_2 k_3} = \frac{9 \epsilon^4}{\pi} \int_0^{2 \pi} \int_0^{2 \pi} \int_0^{2 \pi} \int_0^{2 \pi} \left| \mathbf{L}^{-s -s s s}_{-\mathbf{k} -\mathbf{k}_1 \mathbf{k}_2 \mathbf{k}_3} \right|^2 \delta^{01}_{23}(\mathbf{k}) \mathrm{d}\theta \mathrm{d}\theta_1 \mathrm{d}\theta_2 \mathrm{d}\theta_3.
\label{eq:defition_isotropic_coefficient}
\end{equation}
A subtler point to note here is that the integration in equation (\ref{eq:conservation_N_isotrope}) is constrained by Dirac's delta function $\delta^{01}_{23}(\omega)$ and the definition of the isotropic coefficient in equation (\ref{eq:defition_isotropic_coefficient}). Therefore, it's not just an integration over $(\mathbb{R}^+)^3$, but rather an integration over a more complex manifold. It's important to remember that for gravitational waves in our unit system, $\omega = k$, so $\delta^{01}_{23}(\omega)$ becomes $\delta^{01}_{23}(k)$.

We look for scale-invariant solutions so that, we assume: $N(k) = A k^x$ where $A > 0$ and $x$ are two real constants. We also introduce the dimensionless wavenumber $\xi_i = k_i / k$. Equation (\ref{eq:conservation_N_isotrope}) can thus be written as follow:
\begin{multline}
    \partial_t N(k) = A^3 k^{3x+1} \int_{(\mathbb{R}^+)^3} \mathcal{C}_{1 \xi_1 \xi_2 \xi_3} \left( \xi_1 \xi_2 \xi_3 \right)^x \left( 1 + \xi_1^{\phantom{a}1-x} - \xi_2^{\phantom{a}1-x} - \xi_3^{\phantom{a}1-x} \right) \\[-20pt]
    \times \delta(1 + \xi_1 - \xi_2 - \xi_3) \mathrm{d}\xi_1 \mathrm{d}\xi_2 \mathrm{d}\xi_3.
    \label{eq:conservation_N_isotrope_scaleinvariant}
\end{multline}
Following \citep{galtier_2017}, it is clear that if:
\begin{equation}
	x = 0 \quad \text{or} \quad x = 1,
\end{equation}
the integral vanishes. These solutions correspond to thermodynamic equilibrium in which both the wave action and energy fluxes are zero. Nonetheless, they are not the only accessible solutions. We perform a Zakharov's transform in order to determine other solutions. We will duplicate four times the integral in the right hand side of the equation (\ref{eq:conservation_N_isotrope_scaleinvariant}) and we will perform the following variable changes leaving one of them unchanged:
\begin{subequations}
\begin{align}
    \mathrm{TZa:} \left( \xi_1, \xi_2, \xi_3 \right) &\longrightarrow \left( \frac{1}{\xi_1}, \frac{\xi_2}{\xi_1}, \frac{\xi_3}{\xi_1} \right); \\
    \mathrm{TZb:} \left( \xi_1, \xi_2, \xi_3 \right) &\longrightarrow \left( \frac{\xi_3}{\xi_2}, \frac{1}{\xi_2}, \frac{\xi_1}{\xi_2} \right); \\
    \mathrm{TZc:} \left( \xi_1, \xi_2, \xi_3 \right) &\longrightarrow \left( \frac{\xi_2}{\xi_3}, \frac{\xi_1}{\xi_3}, \frac{1}{\xi_3} \right).
\end{align}
\end{subequations}
We finally have:
\begin{equation}
    \partial_t N(k) = \frac{A^3}{4} k^{3x+1} I(x),
\end{equation}
where:
\begin{multline}
    I(x) = \int_{(\mathbb{R}^+)^3} \mathcal{C}_{1 \xi_1 \xi_2 \xi_3} \left( \xi_1 \xi_2 \xi_3 \right)^x \left( 1 + \xi_1^{\phantom{a}1-x} - \xi_2^{\phantom{a}1-x} - \xi_3^{\phantom{a}1-x} \right) \\[-10pt]
    \times \left( 1 + \xi_1^{\phantom{a}-3x-2} - \xi_2^{\phantom{a}-3x-2} - \xi_3^{\phantom{a}-3x-2} \right) \delta(1 + \xi_1 - \xi_2 - \xi_3) \mathrm{d}\xi_1 \mathrm{d}\xi_2 \mathrm{d}\xi_3.
\end{multline}

Therefore we find two other solutions:
\begin{equation}
    x = -2/3 \quad \text{or} \quad x = -1.
\end{equation}
They have non-zero constant fluxes which indicates the occurrence of cascades. To determine the direction of the cascades, we need to use equations  (\ref{eq:conservationN2}) and (\ref{eq:conservationE2}).
For the case $x = -2/3$, the flux can be expressed as:
\begin{equation}
    \partial_t N(k) = \frac{A^3}{4} k^{3x+1} I(x) = - \partial_k \zeta(k),
\end{equation}
so that:
\begin{equation}
    \zeta(k) = - \frac{A^3}{4}\frac{k^{3x+2}}{3x+2} I(x).
\end{equation}
Therefore, in this case, we have:
\begin{multline}
    \zeta = - \frac{A^3}{4} \int_{(\mathbb{R}^+)^3} \mathcal{C}_{1 \xi_1 \xi_2 \xi_3} \left( \xi_1 \xi_2 \xi_3 \right)^{-2/3} \left( 1 + \xi_1^{5/3} - \xi_2^{5/3} - \xi_3^{5/3} \right) \ln \left( \frac{\xi_2 \xi_3}{\xi_1}\right) \\
    \times \delta \left(1 + \xi_1 - \xi_2 -\xi_3 \right) \mathrm{d}\xi_1 \mathrm{d}\xi_2 \mathrm{d}\xi_3,
\end{multline}
which is negative as a numerical evaluation shows. This indicates the presence of an inverse cascade. Similarly, for the case $x = -1$, we have:
\begin{equation}
\partial_t E(k) = \frac{A^3}{4} k^{3x+2} I(x) = - \partial_k \varepsilon(k)
\end{equation}
so that:
\begin{equation}
\varepsilon(k) = - \frac{A^3}{4} \frac{k^{3x+3}}{3x+3} I(x),
\end{equation}
which, at $x = -1$, yields the energy flow:
\begin{multline}
    \varepsilon = - \frac{A^3}{4} \int_{(\mathbb{R}^+)^3} \mathcal{C}_{1 \xi_1 \xi_2 \xi_3} \left( \xi_1 \xi_2 \xi_3 \right)^{-1} \left( 1 + \xi_1^{2} - \xi_2^{2} - \xi_3^{2} \right) \left( - \xi_1 \ln \xi_1 + \xi_2 \ln \xi_2 + \xi_3 \ln \xi_3 \right) \\
    \times \delta \left(1 + \xi_1 - \xi_2 -\xi_3 \right) \mathrm{d}\xi_1 \mathrm{d}\xi_2 \mathrm{d}\xi_3,
\end{multline}
which is positive as shown by a numerical evaluation, indicating a direct cascade.

In conclusion, the equations of the system (\ref{eq:interest_system}) give rise to two cascades with respective Kolmogorov-Zakharov spectra of the form:
\begin{equation}
N(k) \propto \left( - \zeta \right)^{1/3} k^{-2/3} \quad \text{and} \quad N(k) \propto \varepsilon^{1/3} k^{-1}.
\end{equation}

\section{Conclusion}
\label{sec:conclusion}

Wave turbulence is a successful analytical theory with many applications \cite{nazarenko_2011, galtier_2022}. Our main objective was to use the multiple time scale method originally proposed by Benney and Newell \cite{benney_1966} to derive an integro-differential equation, known as the kinetic equation, for quartic gravity (surface) wave interactions. Unlike the seminal paper, our derivation is very general and assumes only a few symmetries for the interaction coefficient. Consequently, our result can be applied to any problem where nonlinearities are cubic and turbulence statistically homogeneous.

A key aspect of this approach is to identify secular drifts arising from the different decompositions of spectral cumulants. It is these drifts that enable us to establish consistent long-term dynamics and derive the wave kinetic equation. In practice, this requires the evaluation of oscillating integrals via generalized functions theory. 
We have shown that a natural closure is obtained because problematic terms such as sixth-order cumulants are not secular. Therefore, in the long-time limit (which also means $\epsilon \to 0$), these contributions become asymptotically small at main order. Note that this conclusion does not exclude the possibility of sixth-order cumulants contributing to higher orders in the development. 

Furthermore, we have shown that quartic wave turbulence retains a memory up to time $T_3$ in the sense that the evolution of the fourth-order cumulant at zero order depends on the initial condition, so that if it is initially zero, it reaches time $T_3$ without evolving. This property can even be generalized to higher order. 

By applying the multiple time scale method to gravitational waves, we have obtained the kinetic equation that describes the temporal evolution of a set of waves of weak amplitude. The dynamics of gravitational waves is slow, involving a typical time scale proportional to $\tau_{GW}/\epsilon^4$, where  $\tau_{GW} \sim 1 /\omega$ is a linear time and $\epsilon$ is a small parameter measuring the amplitude of gravitational waves.
The kinetic equation asymptotically describes the transfer of wave action and energy through Fourier modes. 
Exact stationary solutions for isotropic turbulence can be obtained using the Zakharov transformation. These solutions are called Kolmogorov-Zakharov spectra. For gravitational wave turbulence, they correspond to a direct cascade of energy and an inverse cascade of wave action. 
As far as we know, these are the first exact solutions of general relativity of a statistical nature.

Our study was initially based on an Eulerian derivation of the wave amplitude equation. This makes a difference with the first study based on a Hamiltonian derivation \cite{galtier_2017}. The kinetic equations obtained are slightly different in their expression, yet retain the same degree of homogeneity. Consequently, the exact solutions are the same, as is the direction of the cascades. The difference is understood as the consequence of an additional symmetry introduced in the Hamiltonian derivation. Both approaches are therefore fully compatible. 

Direct numerical proof of the existence of a dual cascade in gravitational wave turbulence has recently been obtained \cite{galtier_2021}. For the future, it is important to continue this numerical study in order to verify the power law indices and to measure, if possible, the acceleration of the inverse cascade \cite{galtier_2020}. 
Another topic concerns intermittency. In the language of turbulence, this means, on the one hand, checking how far wave amplitudes are from a Gaussian distribution (a small difference is expected) and, on the other, measuring structure functions. The latter involves measuring the difference between a field taken at two positions separated by a distance $L$, all at a given power, and observing the dependence of these functions on $L$. Power laws are expected. These basic fields could be the components of the space-time metric.

\acknowledgments
We acknowledge Sergey Nazarenko for useful discussions. 

\newpage

\appendix
\section{Expression of the six points cummulants}
\label{sec:appendix_six_points_cummulant}

Assuming the means field is equal to zero, the sixth-order moment can be decomposed on the spectral cumulants according to:
\begin{equation}
\begin{aligned}
    \langle a^{s_1}_0(\mathbf{k}_1) a^{s_2}_0(\mathbf{k}_2) a^{s_3}_0(\mathbf{k}_3) a^{s_4}_0(\mathbf{k}_4) a^{s_5}_0(\mathbf{k}_5) a^{s_6}_0(\mathbf{k}_6) \rangle & \\
    & \hspace{-205pt} = q_0^{12}(\mathbf{k}) q_0^{34}(\mathbf{k}) q_0^{56}(\mathbf{k}) + q_0^{12}(\mathbf{k}) q_0^{35}(\mathbf{k}) q_0^{46}(\mathbf{k}) + q_0^{12}(\mathbf{k}) q_0^{36}(\mathbf{k}) q_0^{45}(\mathbf{k}) + q_0^{13}(\mathbf{k}) q_0^{24}(\mathbf{k}) q_0^{56}(\mathbf{k}) \\
    & \hspace{-195pt} + q_0^{13}(\mathbf{k}) q_0^{25}(\mathbf{k}) q_0^{46}(\mathbf{k}) + q_0^{13}(\mathbf{k}) q_0^{26}(\mathbf{k}) q_0^{45}(\mathbf{k}) + q_0^{14}(\mathbf{k}) q_0^{23}(\mathbf{k}) q_0^{56}(\mathbf{k}) + q_0^{14}(\mathbf{k}) q_0^{25}(\mathbf{k}) q_0^{36}(\mathbf{k}) \\
    & \hspace{-195pt} + q_0^{14}(\mathbf{k}) q_0^{26}(\mathbf{k}) q_0^{35}(\mathbf{k}) + q_0^{15}(\mathbf{k}) q_0^{23}(\mathbf{k}) q_0^{46}(\mathbf{k}) + q_0^{15}(\mathbf{k}) q_0^{24}(\mathbf{k}) q_0^{36}(\mathbf{k}) + q_0^{15}(\mathbf{k}) q_0^{26}(\mathbf{k}) q_0^{34}(\mathbf{k}) \\
    & \hspace{-195pt} + q_0^{16}(\mathbf{k}) q_0^{23}(\mathbf{k}) q_0^{45}(\mathbf{k}) + q_0^{16}(\mathbf{k}) q_0^{24}(\mathbf{k}) q_0^{35}(\mathbf{k}) + q_0^{16}(\mathbf{k}) q_0^{25}(\mathbf{k}) q_0^{34}(\mathbf{k}) \\
    & \hspace{-195pt} + q_0^{123}(\mathbf{k}) q_0^{456}(\mathbf{k}) + q_0^{124}(\mathbf{k}) q_0^{356}(\mathbf{k}) + q_0^{125}(\mathbf{k}) q_0^{346}(\mathbf{k}) + q_0^{126}(\mathbf{k}) q_0^{345}(\mathbf{k}) + q_0^{134}(\mathbf{k}) q_0^{256}(\mathbf{k}) \\
    & \hspace{-195pt} + q_0^{135}(\mathbf{k}) q_0^{246}(\mathbf{k}) + q_0^{136}(\mathbf{k}) q_0^{246}(\mathbf{k}) + q_0^{145}(\mathbf{k}) q_0^{236}(\mathbf{k}) + q_0^{146}(\mathbf{k}) q_0^{235}(\mathbf{k}) + q_0^{156}(\mathbf{k}) q_0^{234}(\mathbf{k}) \\
    & \hspace{-195pt} + q_0^{12}(\mathbf{k}) q_0^{3456}(\mathbf{k}) + q_0^{13}(\mathbf{k}) q_0^{2456}(\mathbf{k}) + q_0^{14}(\mathbf{k}) q_0^{2356}(\mathbf{k}) + q_0^{15}(\mathbf{k}) q_0^{2346}(\mathbf{k}) + q_0^{16}(\mathbf{k}) q_0^{2345}(\mathbf{k}) \\
    & \hspace{-195pt} + q_0^{23}(\mathbf{k}) q_0^{1456}(\mathbf{k}) + q_0^{24}(\mathbf{k}) q_0^{1356}(\mathbf{k}) + q_0^{25}(\mathbf{k}) q_0^{1346}(\mathbf{k}) + q_0^{26}(\mathbf{k}) q_0^{1345}(\mathbf{k}) + q_0^{34}(\mathbf{k}) q_0^{1256}(\mathbf{k}) \\
    & \hspace{-195pt} + q_0^{35}(\mathbf{k}) q_0^{1246}(\mathbf{k}) + q_0^{36}(\mathbf{k}) q_0^{1245}(\mathbf{k}) + q_0^{45}(\mathbf{k}) q_0^{1236}(\mathbf{k}) + q_0^{46}(\mathbf{k}) q_0^{1235}(\mathbf{k}) + q_0^{56}(\mathbf{k}) q_0^{1234}(\mathbf{k}) \\
    & \hspace{-195pt} + q_0^{123456}(\mathbf{k}),
\end{aligned}
\end{equation}
where $q_0^{i_1 i_2 \dots i_r}(\mathbf{k}) = q_0^{s_{i_1} s_{i_2} \dots s_{i_r}}(\mathbf{k}_{i_1}, \mathbf{k}_{i_2}, \dots, \mathbf{k}_{i_r}) \delta\left( \sum_{j=1}^r \mathbf{k}_{i_j} \right)$.

\section{Riemann-Lebesgue's lemma for distributions}
\label{sec:appendix_RL_for_distributions}

\noindent
We define for $(x,t) \in \mathbb{R}\times\mathbb{R}^+$, $\Delta(x, t) = \int_0^t e^{ixt'}~\mathrm{d}t' = \frac{e^{ixt}-1}{ix}$. In terms of generalized function, $\Delta(\cdot, t)$ has the following asymptotic behaviour:
\begin{equation}
	\Delta( \cdot , t) \xrightarrow[t \to + \infty]{} \begin{cases}
		\pi \delta( \cdot ) + i \mathcal{P} \left( \frac{1}{\cdot} \right), & \text{if}~x\neq 0; \\
		t, & \text{if}~x = 0. \\
	\end{cases}
\end{equation}

\noindent
\textsc{Proof} - Let $\varphi : x \mapsto \varphi(x)$ be a test function of $C_c^\infty(\mathbb{R})$, this mean that $\varphi$ is infinitely continuously differentiable with a compact support. Thus, we define $M \in \mathbb{R}$ such that:
\begin{equation}
    \forall |x| \geq M,~\varphi(x) = 0.
\end{equation}
We also have:
\begin{equation}
\begin{aligned}
    \langle \Delta(\cdot, t), \varphi \rangle &= \int_\mathbb{R} \frac{e^{ixt} - 1}{ix} \varphi(x)\mathrm{d}x\\
    &= \int_\mathbb{R} \left[\frac{\sin(xt)}{x} \varphi(x) + i \frac{1 - \cos(xt)}{x} \varphi(x) \right]\mathrm{d}x\\
    &= \int_\mathbb{R} \frac{\sin(u)}{u} \varphi\left(\frac{u}{t}\right)\mathrm{d}u + i \int_0^{+\infty} \frac{1 - \cos(xt)}{x} \left[\varphi(x) - \varphi(-x) \right]\mathrm{d}x \\
    &= \int_\mathbb{R} \frac{\sin(u)}{u} \varphi\left(\frac{u}{t}\right)\mathrm{d}u + i \mathcal{P} \left\{ \int_\mathbb{R} \frac{\varphi(x)}{x}\mathrm{d}x \right\} - i \int_0^{+\infty} \frac{\cos(xt)}{x} \left[\varphi(x) - \varphi(-x) \right]~\mathrm{d}x .
\end{aligned}
\end{equation}
Concerning the real part, it is clear that:
\begin{equation}
    \forall x \in \mathbb{R},~\frac{\sin(u)}{u} \varphi\left(\frac{u}{t}\right) \xrightarrow[t \to + \infty]{} \frac{\sin(u)}{u} \varphi\left(0\right),
\end{equation}
and:
\begin{equation}
    \forall x \in \mathbb{R},~\left|\frac{\sin(u)}{u} \varphi\left(\frac{u}{t}\right) \right| \leq 
    \begin{cases}
        0, &\text{ if } |x| \geq M; \\
        \max\limits_{\mathbb{R}} |\varphi|, &\text{ else};
    \end{cases}
\end{equation}
which is integrable. Thus, according to the dominated convergence theorem, the real part converges to:
\begin{equation}
    \varphi(0) \int_\mathbb{R} \frac{\sin(u)}{u}~\mathrm{d}u = \pi \varphi(0).
\end{equation}
Concerning the imaginary part, we may apply the Riemann-Lebesgue's theorem for functions since $v \mapsto \frac{\varphi(v) - \varphi(-v)}{v}$ is integrable upon $\mathbb{R}^+$. Hence,
\begin{equation}
    \int_0^{+\infty} \frac{\cos(vt)}{v} \left[\varphi(v) - \varphi(-v) \right]~\mathrm{d}v \xrightarrow[t \to + \infty]{} 0.
\end{equation}
By gathering all these elements, we have:
\begin{equation}
    \langle \Delta(\cdot, t), \varphi \rangle \xrightarrow[t \to + \infty]{} \pi \varphi(0) + i \mathcal{P} \left\{ \int_\mathbb{R} \frac{\varphi(v)}{v}\mathrm{d}v \right\}.
\end{equation}
Furthermore, it is straightforward to see that if $x=0$, $\Delta(x,t) = t$. $\square$

\section{Consequences of the lemma: other asymptotic properties}
\label{sec:consequences_lemma}

\noindent
We now define for $(x,y,t) \in \mathbb{R}^2\times\mathbb{R}^+$, $E\left(x, y, t\right) = \int_0^t \Delta(x-y,t') e^{iyt'}~\mathrm{d}t'$. In terms of distribution, $E(\cdot, \cdot, t)$ has the following asymptotic behaviour:
\begin{equation}
E(x, y, t) \xrightarrow[t \to + \infty]{} 
\begin{cases}
	\left[\pi \delta(x) + i \mathcal{P} \left( \frac{1}{x} \right)\right] \left[\pi \delta(y) + i \mathcal{P} \left( \frac{1}{y} \right)\right], & \text{if}~x \neq 0,~y \neq 0~\text{and}~x\neq y;\\
	\left[\pi \delta(x) + i \mathcal{P} \left( \frac{1}{x} \right)\right] \left[t - i \frac{\partial}{\partial x}\right], & \text{if}~x \neq 0~\text{and}~y = 0;\\
	i \left[\pi \delta(x) + i \mathcal{P} \left( \frac{1}{x} \right)\right] \frac{\partial}{\partial x}, & \text{if}~ x = y \neq 0;\\
	\frac{t^2}{2}, &\text{if}~x = y = 0. 
\end{cases}
\end{equation}

\noindent
\textsc{Proof} - Let $\varphi : (x,y) \mapsto \varphi(x,y)$ be a test function of $C_c^\infty(\mathbb{R}^2,\mathbb{R})$. \\
\noindent
$\bullet$ We first assume that both $x$ and $y$ are not constrained to vanish and are not constrained to be equal. We have:
\begin{equation}
\begin{aligned}
    \langle E(\cdot, \cdot, t), \varphi \rangle &= \int_{\mathbb{R}^2} \frac{\Delta(x,t) - \Delta(y,t)}{i(x-y)} \varphi(x,y)\mathrm{d}x\mathrm{d}y\\
    &= \int_{\mathbb{R}} \Delta(x,t) \mathcal{P}\left\{ \int_{\mathbb{R}} \frac{\varphi(x,y)}{i(x-y)} \mathrm{d}y\right\} \mathrm{d}x - \int_{\mathbb{R}} \Delta(y,t) \mathcal{P} \left\{ \int_{\mathbb{R}} \frac{\varphi(x,y)}{i(x-y)} \mathrm{d}x\right\} \mathrm{d}y .\\
\end{aligned}
\end{equation}
Thus, when taking the limit $t \to +\infty$, we obtain:
\begin{equation}
\begin{aligned}
    \langle E(\cdot, \cdot, t), \varphi \rangle \xrightarrow[t \to + \infty]{}& ~ i \pi \mathcal{P}\left\{ \int_{\mathbb{R}} \frac{\varphi(0,y)}{y} \mathrm{d}y \right\} + \mathcal{P} \left\{ \int_\mathbb{R} \frac{1}{x} \mathcal{P}\left\{\int_{\mathbb{R}} \frac{\varphi(x,y)}{(x-y)} \mathrm{d}y \right\} \mathrm{d}x \right\} \\ 
    & + i \pi \mathcal{P}\left\{ \int_{\mathbb{R}} \frac{\varphi(x,0)}{x} \mathrm{d}x \right\} + \mathcal{P} \left\{ \int_\mathbb{R} \frac{1}{y} \mathcal{P}\left\{ \int_{\mathbb{R}} \frac{\varphi(x,y)}{(x-y)} \mathrm{d}x \right\} \mathrm{d}y \right\}. \\
\end{aligned}
\end{equation}
Hence, when using the Poincaré--Bertrand lemma, we have:
\begin{multline}
    \mathcal{P} \left\{ \int_\mathbb{R} \frac{1}{x} \mathcal{P}\left\{\int_{\mathbb{R}} \frac{\varphi(x,y)}{(x-y)} \mathrm{d}y \right\} \mathrm{d}x \right\} + \mathcal{P} \left\{ \int_\mathbb{R} \frac{1}{y} \mathcal{P}\left\{ \int_{\mathbb{R}} \frac{\varphi(x,y)}{(x-y)} \mathrm{d}x \right\} \mathrm{d}y \right\} \\
    = - \pi^2 \varphi(0,0) + \mathcal{P} \left\{ \int_{\mathbb{R}^2} \frac{\varphi(x,y)}{xy} \mathrm{d}x\mathrm{d}y \right\}.
\end{multline}
This result allows us to write:
\begin{multline}
    \langle E(\cdot, \cdot, t), \varphi \rangle \xrightarrow[t \to + \infty]{} ~ - \pi^2 \varphi(0,0) + \mathcal{P} \left\{ \int_{\mathbb{R}^2} \frac{\varphi(x,y)}{xy} \mathrm{d}x\mathrm{d}y \right\} \\
    + i \pi \mathcal{P}\left\{ \int_{\mathbb{R}} \frac{\varphi(0,y)}{y} \mathrm{d}y \right\} + i \pi \mathcal{P}\left\{ \int_{\mathbb{R}} \frac{\varphi(x,0)}{x} \mathrm{d}x \right\}.
\end{multline}
Finally, if $x \neq 0$, $y \neq 0$ and $x \neq y$, we have:
\begin{equation}
    E(\cdot, \cdot, t) \xrightarrow[t \to + \infty]{} \left[\pi \delta(x) + i \mathcal{P} \left( \frac{1}{x} \right)\right] \left[\pi \delta(y) + i \mathcal{P} \left( \frac{1}{y} \right)\right].
\end{equation}
$\bullet$ Now, we assume that only $y$ is constrained to be equal to zero. In this case, we have:
\begin{equation}
\begin{aligned}
    \langle E(\cdot, 0, t), \varphi \rangle &= \int_{\mathbb{R}} \frac{e^{ixt} - 1 - ixt}{(ix)^2} \varphi(x)\mathrm{d}x\\
    &= \int_{\mathbb{R}} - \frac{1}{x} \left[ e^{ixt} \frac{\mathrm{d}\varphi}{\mathrm{d}x}(x) + i t \varphi(x) e^{ixt} \right] + \frac{1}{x} \frac{\mathrm{d}\varphi}{\mathrm{d}x}(x) - \frac{t}{ix} \varphi(x) \mathrm{d}x \\
    &= \int_\mathbb{R} \Delta(x,t) \left(t - i \frac{\mathrm{d}}{\mathrm{d}x} \right) \varphi(x) \mathrm{d}x.
\end{aligned}
\end{equation}
Thus, when taking the limit $t \to +\infty$ we find the asymptotic behaviour:
\begin{equation}
    E(\cdot, 0, t) \xrightarrow[t \to + \infty]{} \left[\pi \delta(x) + i \mathcal{P} \left( \frac{1}{x} \right)\right] \left(t - i \frac{\mathrm{d}}{\mathrm{d}x} \right).
\end{equation}
$\bullet$ When $y$ is equal to $x$, we may notice that:
\begin{equation}
    E(x,x,t) = t \Delta(x,t) - E(x,0,t).
\end{equation}
Thus, the asymptotic behaviour is given by:
\begin{equation}
    E(x, x, t) \xrightarrow[t \to + \infty]{} i \left[\pi \delta(x) + i \mathcal{P} \left( \frac{1}{x} \right)\right] \frac{\mathrm{d}}{\mathrm{d}x} .
\end{equation}
$\bullet$ Last but not least, it is clear that if $x = y = 0$, we have $E(\cdot, \cdot, t) = t^2 / 2$. $\square$

\vspace{1\baselineskip}

\noindent
In terms of generalized function, $\Delta(x, t) \Delta(-x, t)$ has the following asymptotic behaviour:
\begin{equation}
\Delta(x, t) \Delta(-x, t) \xrightarrow[t \to + \infty]{} 2 \pi t \delta(x) + 2 \mathcal{P} \left( \frac{1}{x} \right) \frac{\mathrm{d}}{\mathrm{d}x} .
\end{equation}

\noindent
\textsc{Proof} - We may notice that:
\begin{equation}
    \Delta(x,t) \Delta(-x,t) = E(x,0,t) + E(-x,0,t).
\end{equation}
Yet, we have:
\begin{equation}
    E(-x,0,t) \xrightarrow{t \to + \infty} \left[\pi \delta(x) - i \mathcal{P} \left( \frac{1}{x} \right)\right] \left(t + i \frac{\mathrm{d}}{\mathrm{d}x} \right).
\end{equation}
Hence, the asymptotic behaviour is given by:
\begin{equation}
    \Delta(x,t) \Delta(-x,t) \xrightarrow{t \to + \infty} 2 \pi t \delta(x) + 2 \mathcal{P}\left( \frac{1}{x} \right) \frac{\mathrm{d}}{\mathrm{d}x}. \quad \square
\end{equation}

\end{document}